\documentclass[
 superscriptaddress,
 reprint,
 amsmath,amssymb,
 aps,
 prb,
]{revtex4-2}

\usepackage{graphicx}
\usepackage{dcolumn}
\usepackage{bm}
\usepackage{subfigure}
\usepackage{graphicx}  
\usepackage{color}

\begin{document}
 \newcommand{\comment}[1]{{\color{red}#1}}

 \newcommand{\Csix}[0]{(\( C_2 \parallel C_{6z}t \))}
 \newcommand{\Mz}{\ensuremath{(C_2 \parallel \widetilde{M}_{(001)})}}
\newcommand{\Mxy}{\ensuremath{(C_2 \parallel \widetilde{M}_{(1\overline{1}0)})}}

\preprint{APS/123-QED}

\title{Altermagnetism of ultrathin CrSb slabs}

\newcommand{\ifm}{Theoretical Physics Division, Department of Physics, Chemistry and Biology (IFM), Link\"oping University, SE-581 83 Link\"oping, Sweden}

\author{Brahim Marfoua} \email {brahim.marfoua@liu.se}
\affiliation{\ifm}

\author{Mohammad Amirabbasi} 
\affiliation{Technische Universit\"at Darmstadt, Fachbereich Material und Geowissenschaften, Fachgebiet Materialmodellierung, Otto‑Berndt‑Straße 3, 6428 Darmstadt, Germany}
\author{Marcus Ekholm}\email {marcus.ekholm@liu.se}
\affiliation{\ifm}

\begin{abstract}
Altermagnets exhibit momentum-dependent spin splitting without net magnetization, combining characteristics of both ferromagnets and antiferromagnets, making them highly interesting for spintronics applications. CrSb is a prime candidate with a high N\'eel temperature ($\sim700$~K) and a large exchange-driven splitting of $\sim0.6$--1~eV. Using ab-initio calculations, we consider slabs of various orientations in the ultrathin limit. In (0001) oriented slabs, the exchange-driven altermagnetic spin splitting collapses, but including spin-orbit coupling restores a residual anisotropic splitting of $\sim70$~meV. The (100)-oriented slabs become fully spin-degenerate due to symmetry reduction. In contrast, the (110) oriented slabs show a strong altermagnetic spin splitting of $\sim400$~meV, and thus emerges as a robust candidate for realizing large, exchange-driven altermagnetism.
\end{abstract}

\maketitle

\section{Introduction}
Altermagnetism has recently emerged as a novel magnetic order, combining features of both ferromagnetism and antiferromagnetism \cite{Mazin2022, Smejkal2022, Lee2024, Krempasky2024, Feng2022}. 
Like ferromagnets (FMs), altermagnets (AMs) exhibit spin splitting of the electronic bands, yet, akin to antiferromagnets (AFMs), they retain zero net magnetization. 

Unlike conventional collinear AFMs in the non-relativistic limit, the opposite-spin sublattices are not related by a lattice translation ($t$) or inversion ($i$), but by a rotation (proper or improper, symmorphic or nonsymmorphic) \cite{Smejkal2022a}.
The symmetry of AMs is thus characterized by a non-relativistic spin-rotation operation, $R^S_i$, and a real-space transformation, $R_j$, combined in a so-called spin-group operator, $\left( R^S_i \parallel R_j  \right)$. Time-reversal symmetry ($\mathcal{T}$) is thus broken, but Kramers degeneracy in AFMs, $E(\mathbf{k}, \uparrow) = E(\mathbf{k}, \downarrow)$, is not restored \cite{Brinkman1966,Smejkal2022a}. The unique band structure of AMs may enable these properties to be harnessed simultaneously in a single material without external stimuli. This intrinsic ability could lead to more efficient spintronic devices, relying on the material’s properties rather than complex device architectures or external fields.

Numerous potential altermagnetic materials have been proposed theoretically, with experimental verification already achieved for several candidates, notably RuO$_2$, MnTe and CrSb \cite{Fedchenko2024, Guo2024, Osumi2024, Hariki2024}. In particular, CrSb has received increasing attention due to its high N\'eel temperature (705 K) \cite{takei63} and substantial spin splitting of 0.6--1 eV reported by ARPES measurements \cite{Yang2024, Li2024, Ding2024, Lu2024, Reimers2024, rai2025direction, Zhou2025, zhang2025chiral, yu2025neel}.
In transport, CrSb exhibits high carrier mobility, with nonlinear Hall resistivity at low temperatures attributed to a multi-carrier effect rather than an anomalous Hall effect \cite{Urata2024}, a property advantageous for efficient spin-current generation.
At higher temperatures, manipulation of the altermagnetic order via crystal symmetry has been demonstrated, showing control of the N\'eel vector orientation and giving rise to a spontaneous anomalous Hall effect at room temperature, which may be important for magnonics applications \cite{Zhou2025}.

\begin{figure}
\vspace{1.05cm}
    \centering
    \hspace*{-0.7cm}
    \includegraphics[width=1.1\linewidth]{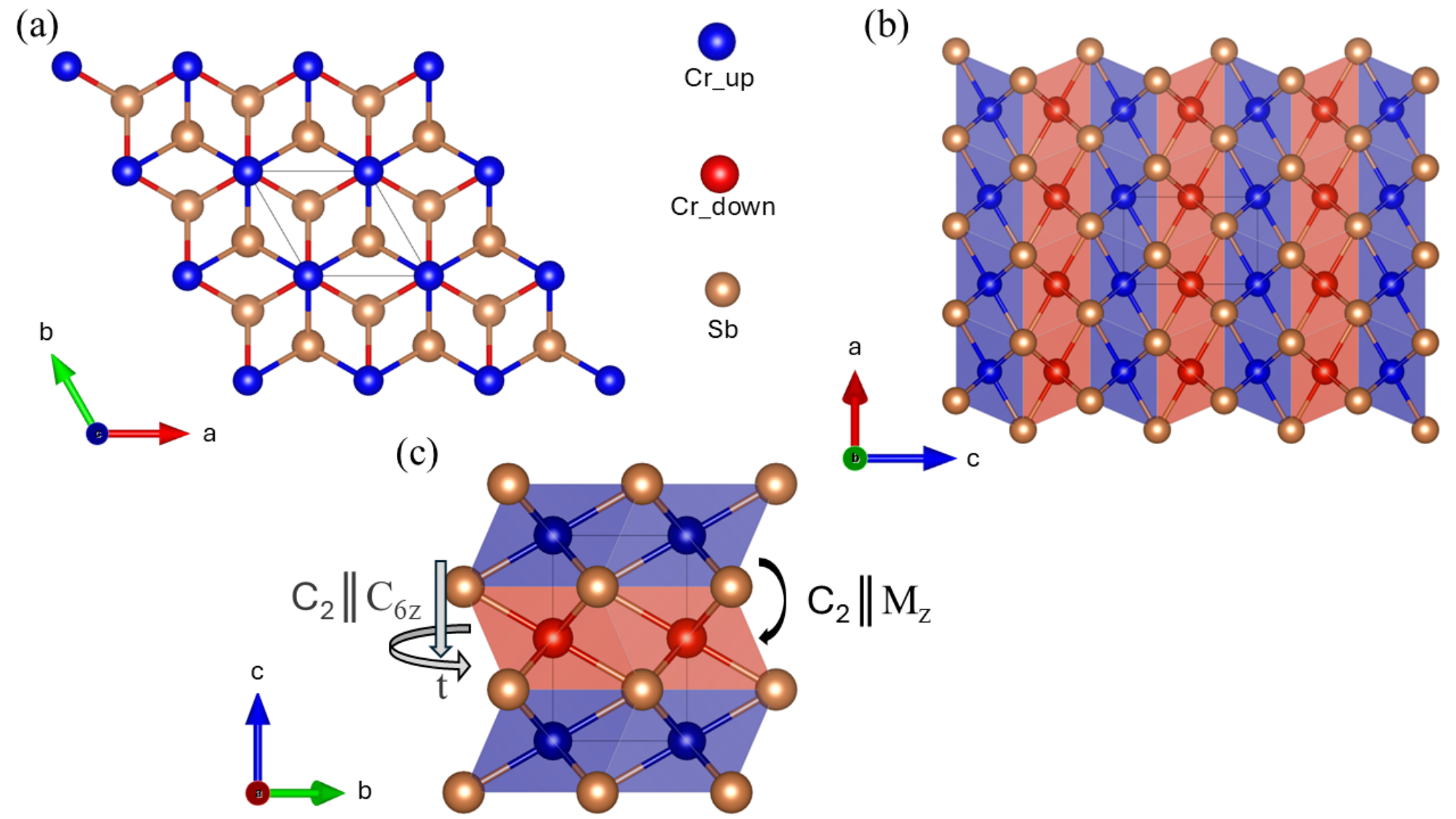}
    \caption{Structural illustration of the hexagonal CrSb bulk system along the (a) $ab$-plane and (b) $ac$-plane. (c) Illustration of two symmetry operations connecting the opposite spin sub-lattices in the CrSb system. }
    \label{fig:struct}
\end{figure}

CrSb crystallizes in a hexagonal NiAs-type structure, with space group $P6_3/mmc$ (No.\ 194), and lattice parameters \( a = b=4.103 \, \text{\AA}\) and \( c = 5.463 \ \text{\AA} \) \cite{venkatraman1990}, 
as shown in Fig.~\ref{fig:struct}. 
Neutron diffraction has revealed collinear antiparallel order along the $c$-axis \cite{Urata2024}.
The spin-group operations protecting the AM splitting consist of a spin flip followed by a screw operation, \Csix, and a spin flip combined with a glide operation such as \Mz\ and \Mxy, as illustrated in Fig.~\ref{fig:struct}(c).
In addition to the horizontal glide, the $C_{6z}$ rotation generates three symmetry-equivalent vertical glides, giving a total of four independent spin-group operations that connect the opposite-spin sub-lattices.

Recent theoretical work has also clarified the conditions under which antisymmetric exchange may arise in altermagnets. In particular, Sim \textit{et al.}~\cite{Sim2025} showed that the Dzyaloshinskii-Moriya interaction (DMI) in CrSb is forbidden in the unstrained crystal, as the Cr--Cr exchange paths are related by inversion and mirror symmetries that suppress the DM vectors according to Moriya's rules. Their analysis further demonstrated that a finite, staggered DMI can appear when external strain breaks the threefold rotational symmetry of the lattice. Therefore, and in the absence of such strain, DMI is expected to be negligible in the CrSb.

CrSb films have recently been grown as films of different thicknesses and orientations \cite{Reimers2024,Ding2024,santhosh2025, aota2025}.
While the altermagnetic features of bulk CrSb are well established, it remains uncertain whether altermagnetic properties are preserved as the material approaches the single-nanometer regime \cite{santhosh2025}. This question is crucial for assessing the feasibility of realizing ultrathin altermagnetic devices.

In this study, we explore various exfoliation planes in CrSb via \textit{ab initio} calculations to identify those capable of exhibiting altermagnetic properties in the extreme, ultrathin limit. 
In particular, we focus on AFM CrSb slabs with (0001), (100) and (110) orientations. We find that, unlike the (0001) and (100) slabs, the (110) oriented slab shows a robust AM splitting down to the width of a single unit cell.
We also show that the strong magnetic exchange-couplings of the bulk are augmented in (110)-oriented ultrathin films, making them highly interesting for future applications in spintronic devices, where the ability to sustain spin splitting at reduced dimensionality is essential.

\section{Numerical Methods}

Spin-polarized first-principles calculations were carried out using the Vienna \textit{Ab initio} Simulation Package (VASP) within the projector augmented-wave (PAW) framework \cite{Kresse1999, Blochl1994}. The exchange–correlation energy was described using the generalized gradient approximation (GGA) in the Perdew–Burke–Ernzerhof (PBE) parametrization \cite{Perdew1996}. For the valence configurations, Cr included the $3p$, $3d$, and $4s$ states in the valence, while Sb included the $5s$ and $5p$ electrons. The PBE functional was chosen because it reliably reproduces experimental structural and electronic properties of CrSb \cite{Reimers2024}. 

A plane-wave energy cutoff of 650~eV was applied for all calculations. The total-energy convergence criterion was set to $10^{-6}$~eV, and the atomic positions were relaxed until the Hellmann–Feynman forces were below $0.01$~eV/\AA. For slab models, a vacuum region of 25~\AA\ was introduced along the out-of-plane direction to prevent spurious interactions between periodic images. The Brillouin zone was sampled using a Monkhorst–Pack $k$-point mesh of $13 \times 15 \times 7$ for bulk calculations and $13 \times 15 \times 1$ for monolayers and slabs.

To evaluate magnetic exchange interactions ($J$), we constructed a tight-binding Hamiltonian from maximally localized Wannier functions (MLWFs) using the \textsc{Wannier90} package \cite{Pizzi2020}. The $J$ parameters were then extracted using the Green’s function formalism within the magnetic force theorem, as implemented in the \textsc{TB2J} code \cite{He2021}.

\section{Results}
\subsection{CrSb bulk system}
\begin{figure}
    \centering
    \includegraphics[width=1\linewidth]{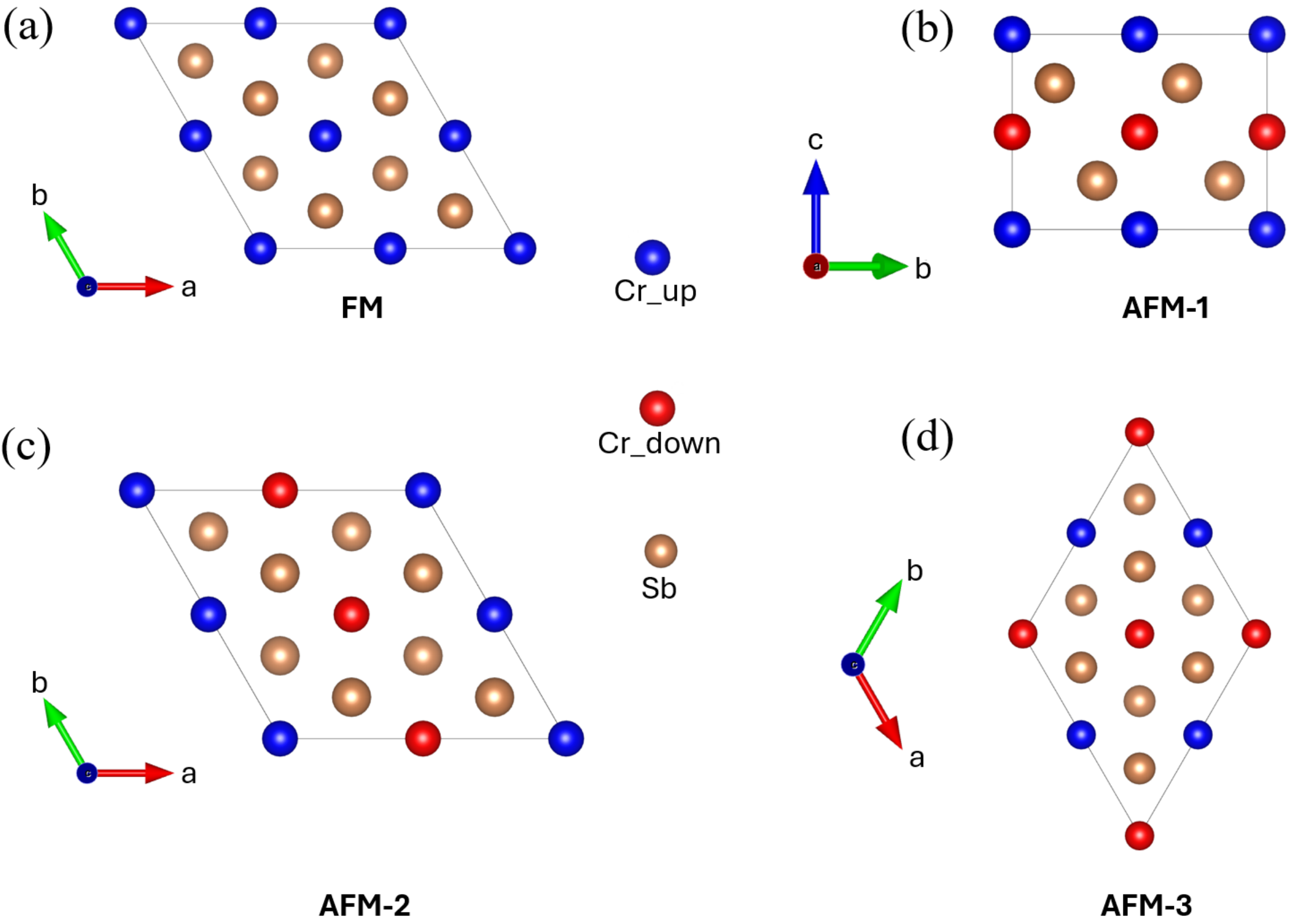}
    \caption{Illustration of different considered magnetic configurations: (a) FM, (b) AFM-1 (layered (001)), (c) AFM-2 (layered (100)), and (d) AFM-3 (layered (110)). Blue/red spheres denote spin-up/down Cr atoms.}
    \label{fig:afm-config}
\end{figure}

To validate our computational framework, and set a baseline for the slab calculations, we first consider bulk properties of $\mathrm{CrSb}$ using experimental lattice parameters. We compare the relative energies of the FM state, and the collinear AFM-1, AFM-2, and AFM-3 configurations illustrated in Fig.~\ref{fig:afm-config}. As summarized in Table~\ref{tab:Ediff-bulk}, the AFM-1 N\'eel state, with alternating spins along the $c$-axis, is lowest in energy. The local magnetic moment of the Cr atoms is $2.79 \mu_B$, which is close to the reported experimental values of $2.7\pm0.2 \mu_B$ \cite{snow52} or $2.84\mu_B$ at room temperature, which was extrapolated to $3\mu_B$ at 0 K \cite{takei63}. The FM state is 81.82~meV/atom higher, while the AFM-2 and AFM-3 states are $\sim$ 86~meV/atom above the ground state. Magnetic anisotropy calculations yield an out-of-plane easy axis with an energy of 0.2~meV/atom, consistent with prior reports~\cite{Reimers2024}.

\begin{table}
\centering
\caption{Energy differences ($\Delta E$) relative to the AFM (001) state, and local Cr magnetic moments for various configuration in bulk CrSb.}
\label{tab:Ediff-bulk}
\begin{ruledtabular}
\begin{tabular}{lcccc}
CrSb-bulk & FM & AFM-1 & AFM-2 & AFM-3 \\
\hline
$\Delta E$  (meV/atom) & 81.82 & 0.00 & 86.62 & 86.25 \\
$m_\mathrm{Cr}$ ($\mu_B$) & 2.64 & 2.79 & 2.50 & 2.50\\
\end{tabular}
\end{ruledtabular}
\end{table}

\begin{figure*}[hbt!]
    \centering
    \subfigure[\label{fig:BZ}]{\includegraphics[width=0.45\linewidth]{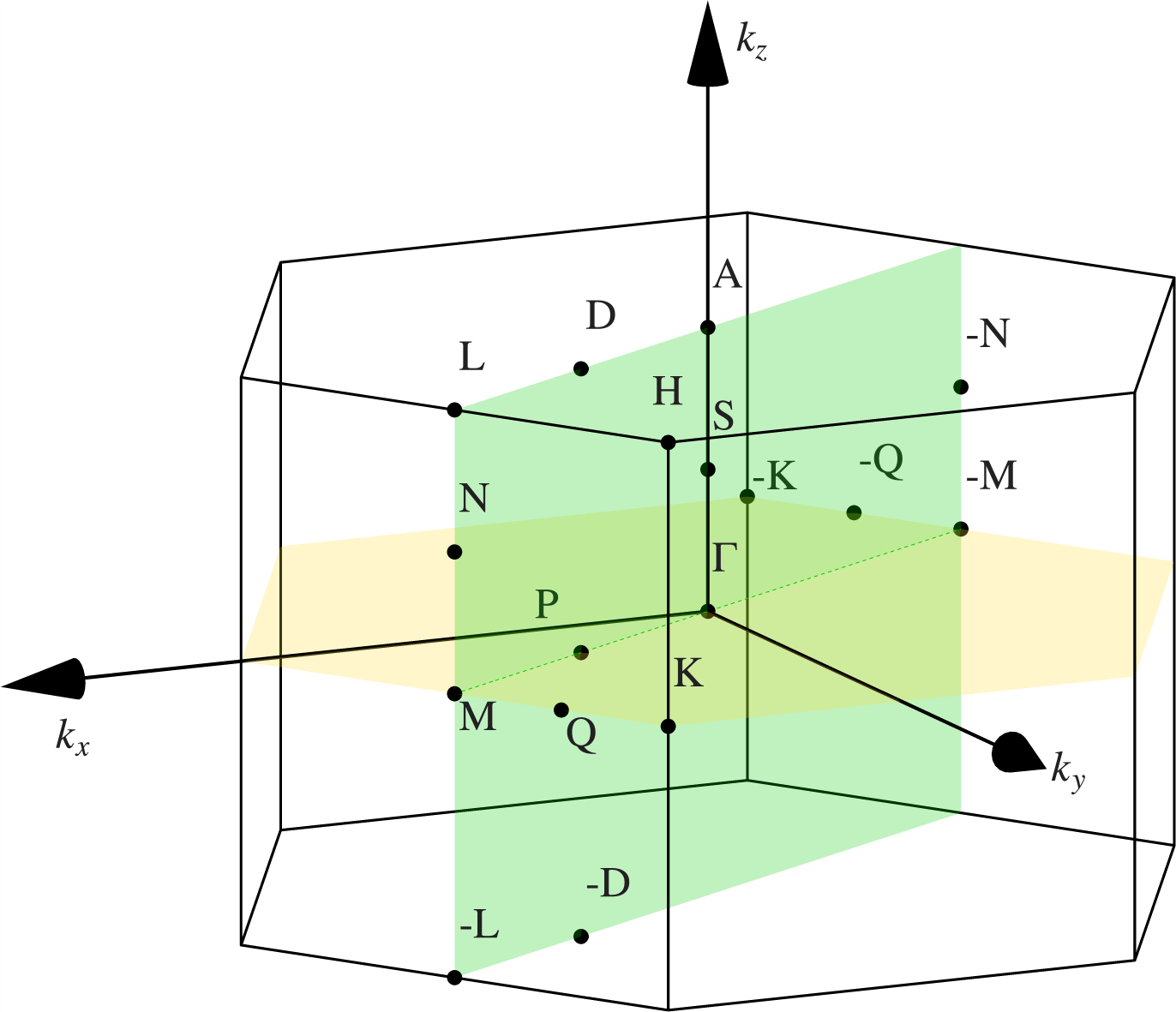}}
    \subfigure[\label{fig:band-1-HS}]{\includegraphics[width=0.49\linewidth]{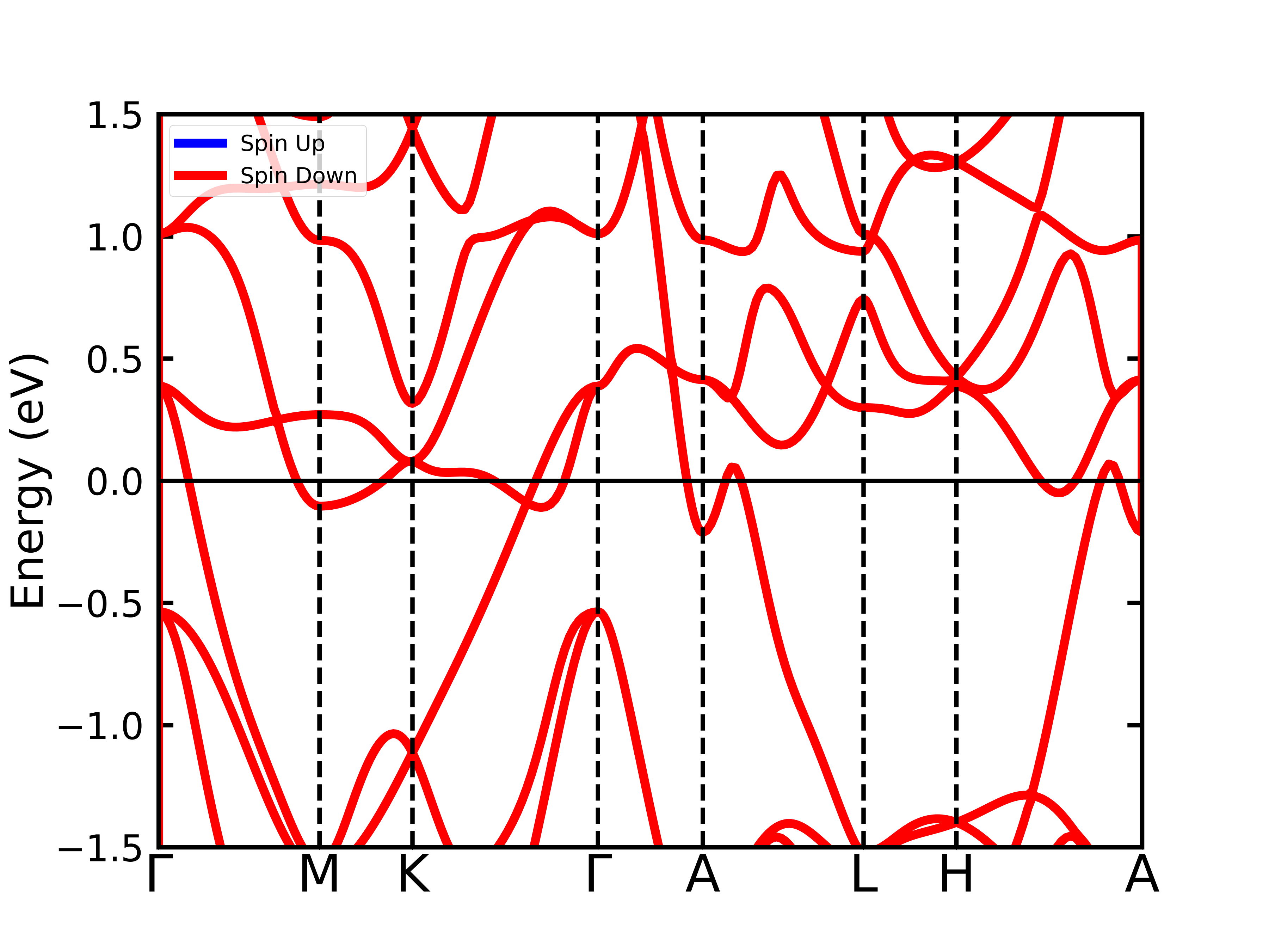}}\hfill
    \subfigure[\label{fig:band-1-LS}]{\includegraphics[width=0.49\linewidth]{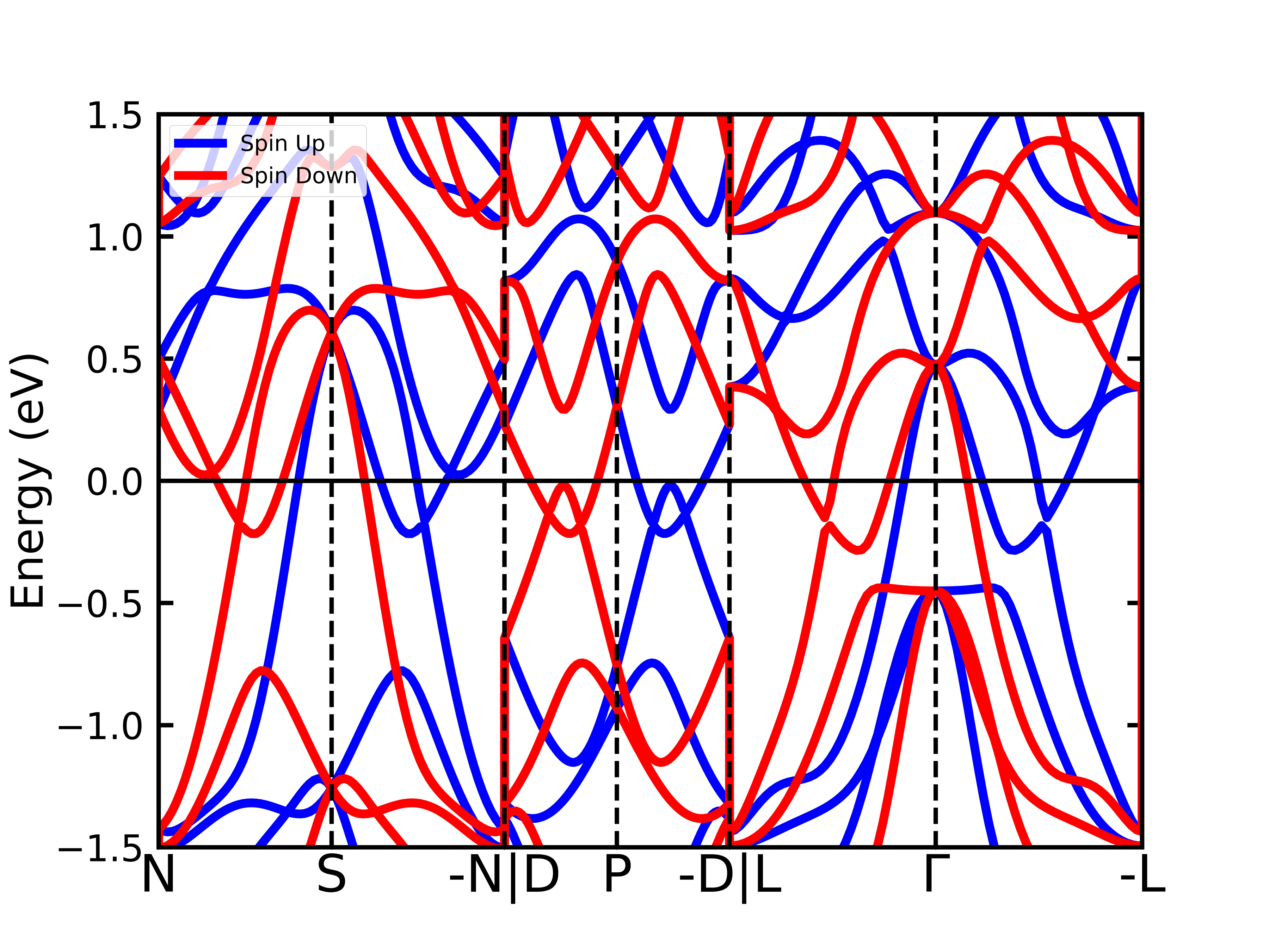}} 
    \subfigure[\label{fig:band-1-LS-soc}]{\includegraphics[width=0.49\linewidth]{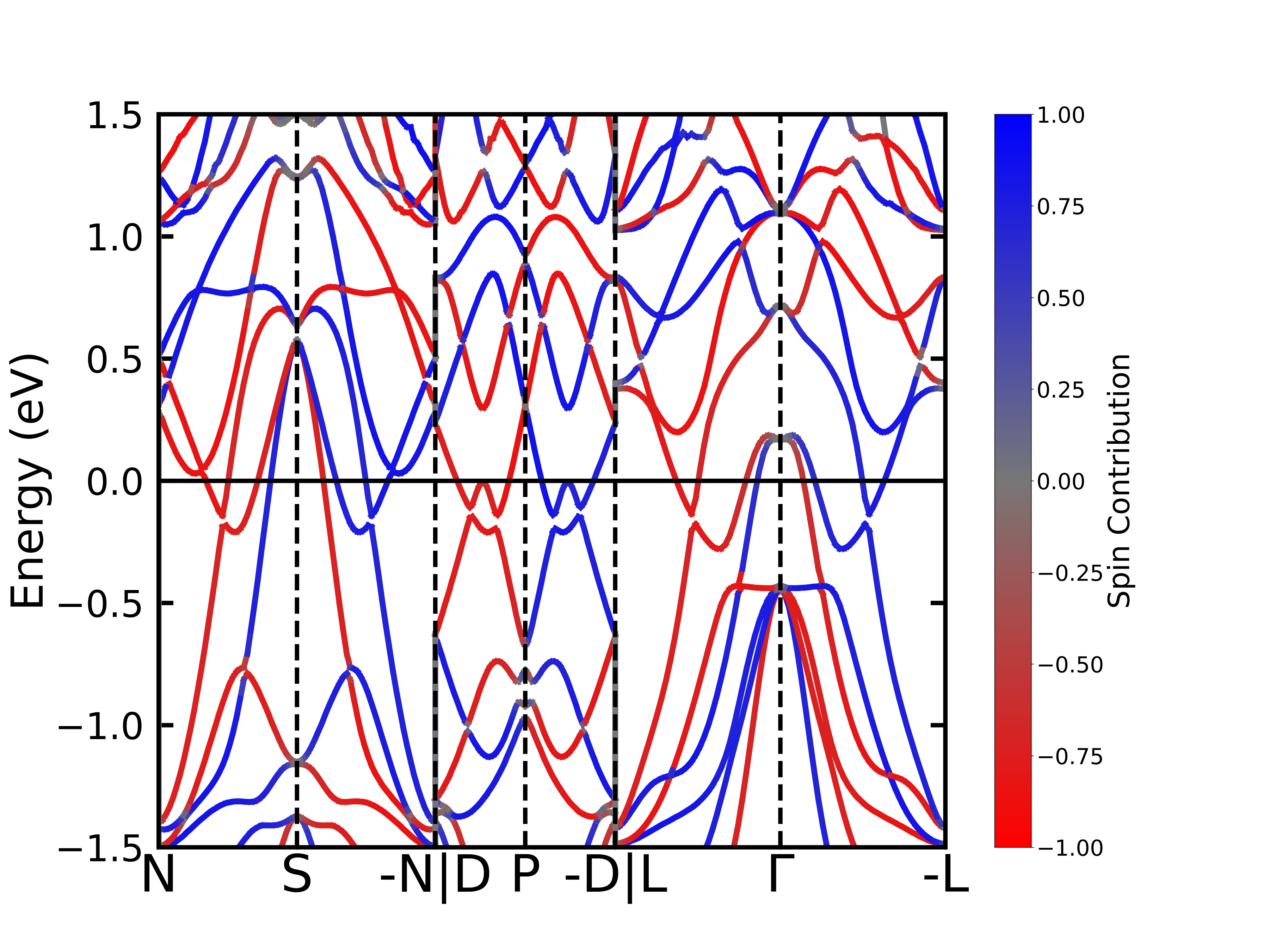}}
     \caption{(a) First BZ of hexagonal bulk CrSb. (b) Scalar-relativistic band structure of the bulk along high-symmetry lines, and (c) along low-symmetry lines. (d) Band structure including SOC along low-symmetry lines. 
     }
    \label{fig:dos-bulk}
\end{figure*}
Fig.~\ref{fig:dos-bulk}(a) shows the first Brillouin zone (BZ) of hexagonal bulk $\mathrm{CrSb}$, which exhibits a metallic electronic structure.
The band structure along the high-symmetry path $\Gamma$–M–K–$\Gamma$, shown in Fig.\ \ref{fig:band-1-HS}, is spin-degenerate and consistent with that of a conventional compensated AFM metal \cite{Reimers2024, Ding2024}. 
This is because the antiunitary $\mathcal{T}\cdot i$ symmetry enforces $E(\mathbf{k},\uparrow) = E(\mathbf{k},\downarrow)$ along those paths.
In contrast, along low-symmetry paths, shown in Fig.~\ref{fig:band-1-LS}, we find pronounced momentum-dependent spin splitting of $0.6$--$1.0$~eV near $E_F$.
Here, $\mathcal{T}\cdot i$ is broken, so that $E(\mathbf{k},\uparrow) \neq E(\mathbf{k},\downarrow)$, although the combined relation $E(\mathbf{k},\uparrow) = E(-\mathbf{k},\downarrow)$ still holds, ensuring zero net moment. This is the hallmark of bulk (B-4) $g$-wave altermagnetism, where the spin splitting in momentum space produces a sixfold alternating pattern of spin polarization across the Brillouin zone in CrSb bulk

\begin{figure*}
    \centering
    \subfigure[]{\includegraphics[width=0.49\textwidth]{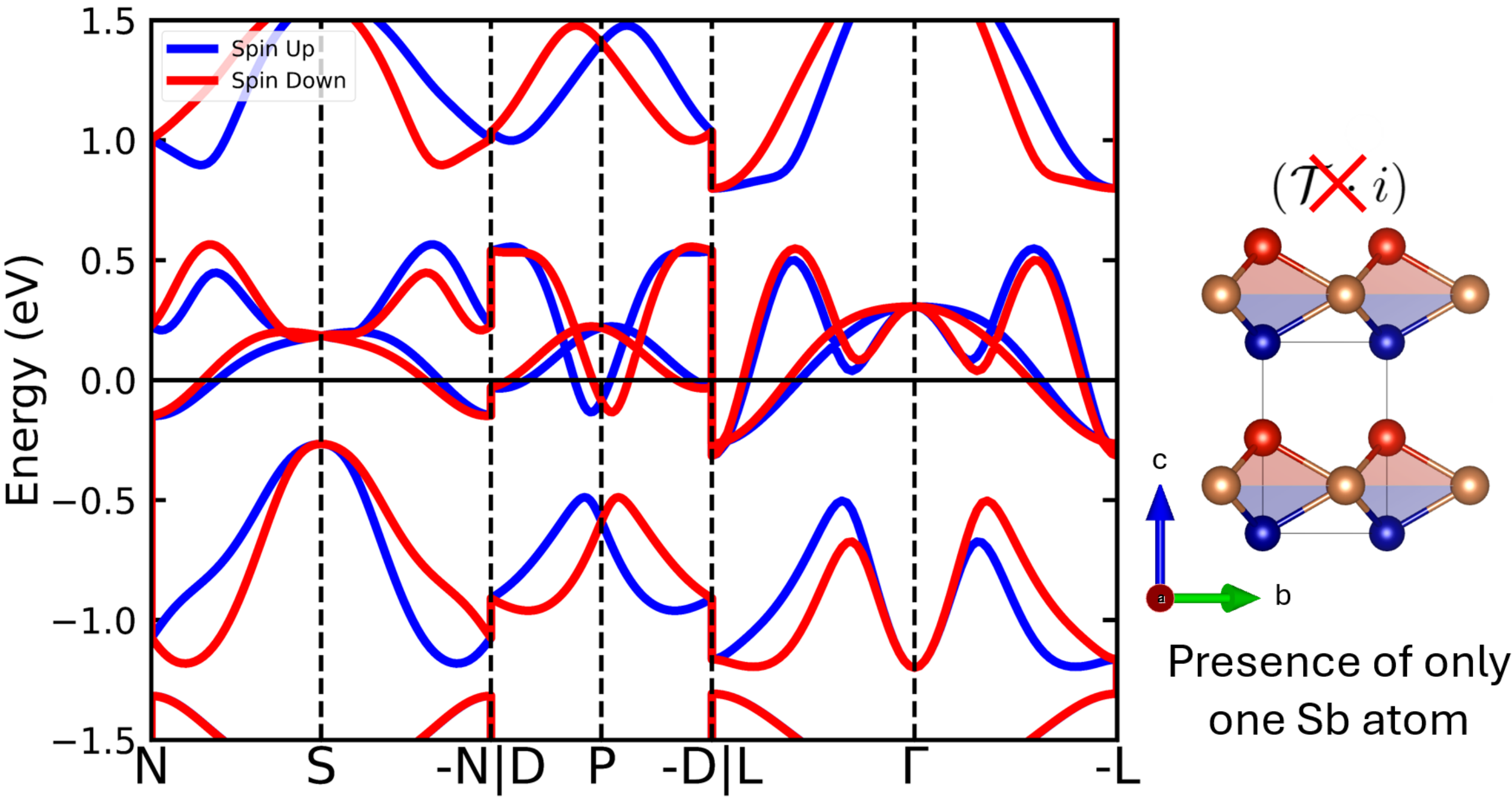}}
    \subfigure[]{\includegraphics[width=0.49\textwidth]{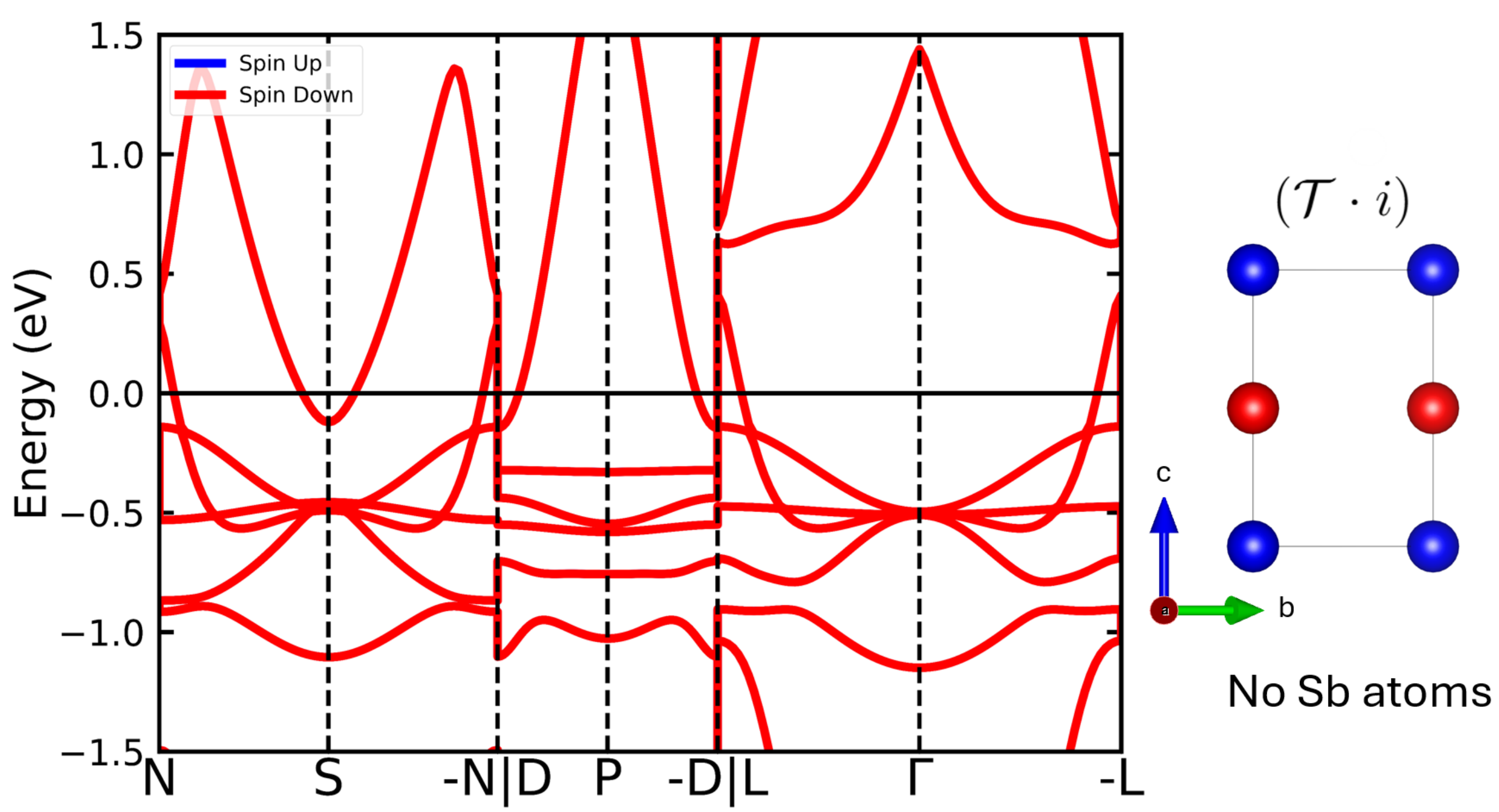}}
    \caption{Spin-polarized bands along low-symmetry paths for (a) Cr$_2$Sb and (b) hcp-Cr. Spin-up and spin-down atoms are shown in blue and red.}
    \label{fig:no-sb}
\end{figure*}

Including spin-orbit coupling (SOC), as shown in Fig.~\ref{fig:band-1-LS-soc}, does not influence the AM splitting, confirming that the large splitting is purely exchange-driven and protected by the bulk spin-group symmetry.
To elaborate on this point we have computed the band structure while removing Sb atoms. Removing the Sb-atoms of every other $ab$-plane, we obtain the artificial Cr$_2$Sb system shown in Fig.~\ref{fig:no-sb}(a). While the AM splitting is reduced, it still persists, as the spin-group operations are preserved, with a large local Cr magnetic moment of $3.5~\mu_\mathrm{B}$. When all Sb atoms are removed, we obtain hcp-structured Cr, which is a conventional AFM with a local magnetic moment of $4~\mu_\mathrm{B}$. Thus, Sb atoms are essential for breaking the $\mathcal{T} \cdot i$ symmetry and enabling the exchange-driven $g$-wave spin texture.

\subsection{(0001) slab}\label{sec:slab0001}
\begin{figure*}[hbt!]
    \centering
    \subfigure[\label{fig:001-illustration}]{\includegraphics[trim=0 0 0 50,clip,width=0.68\linewidth]{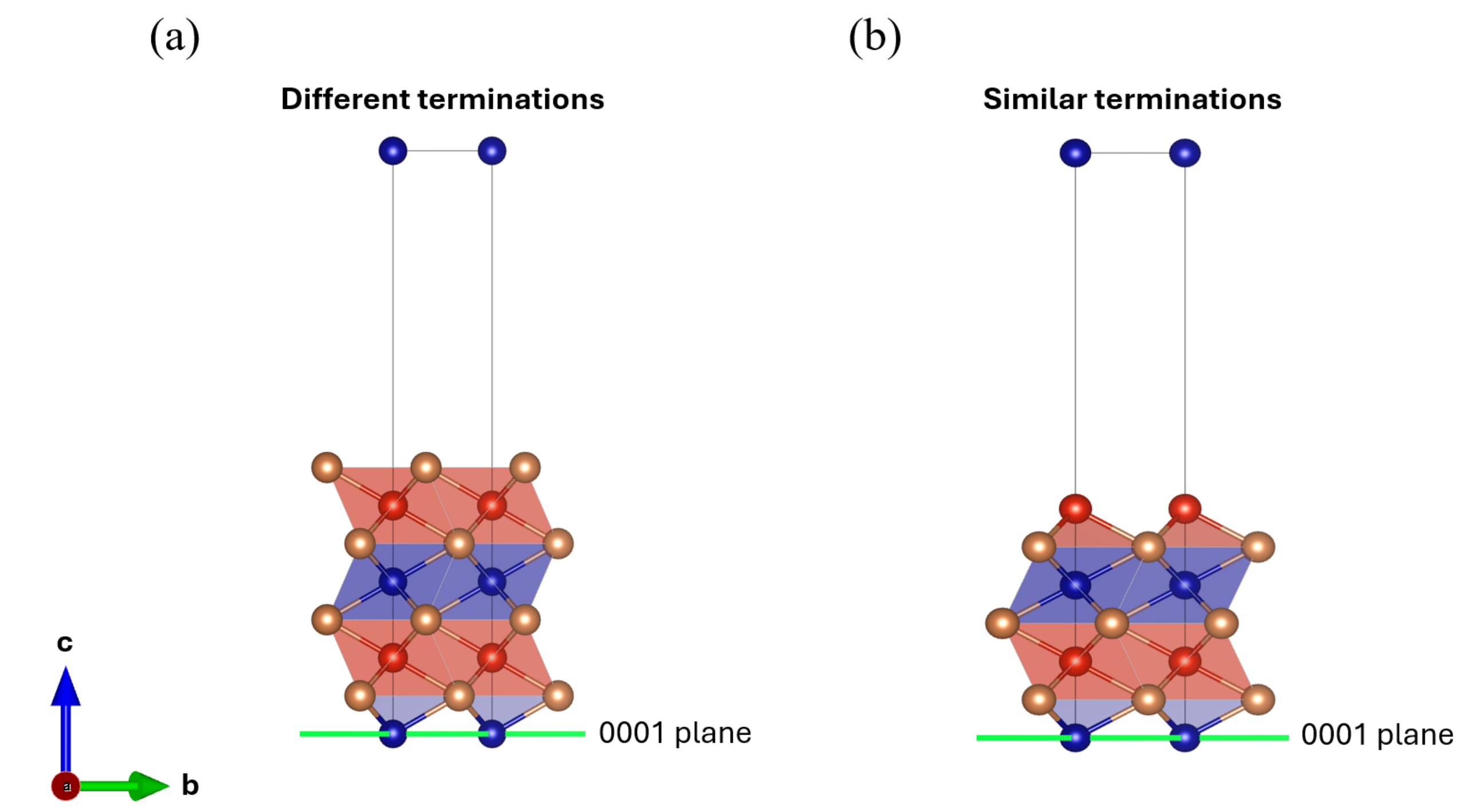}}
    \subfigure[\label{fig:001-noSOC}]{\includegraphics[width=0.49\linewidth]{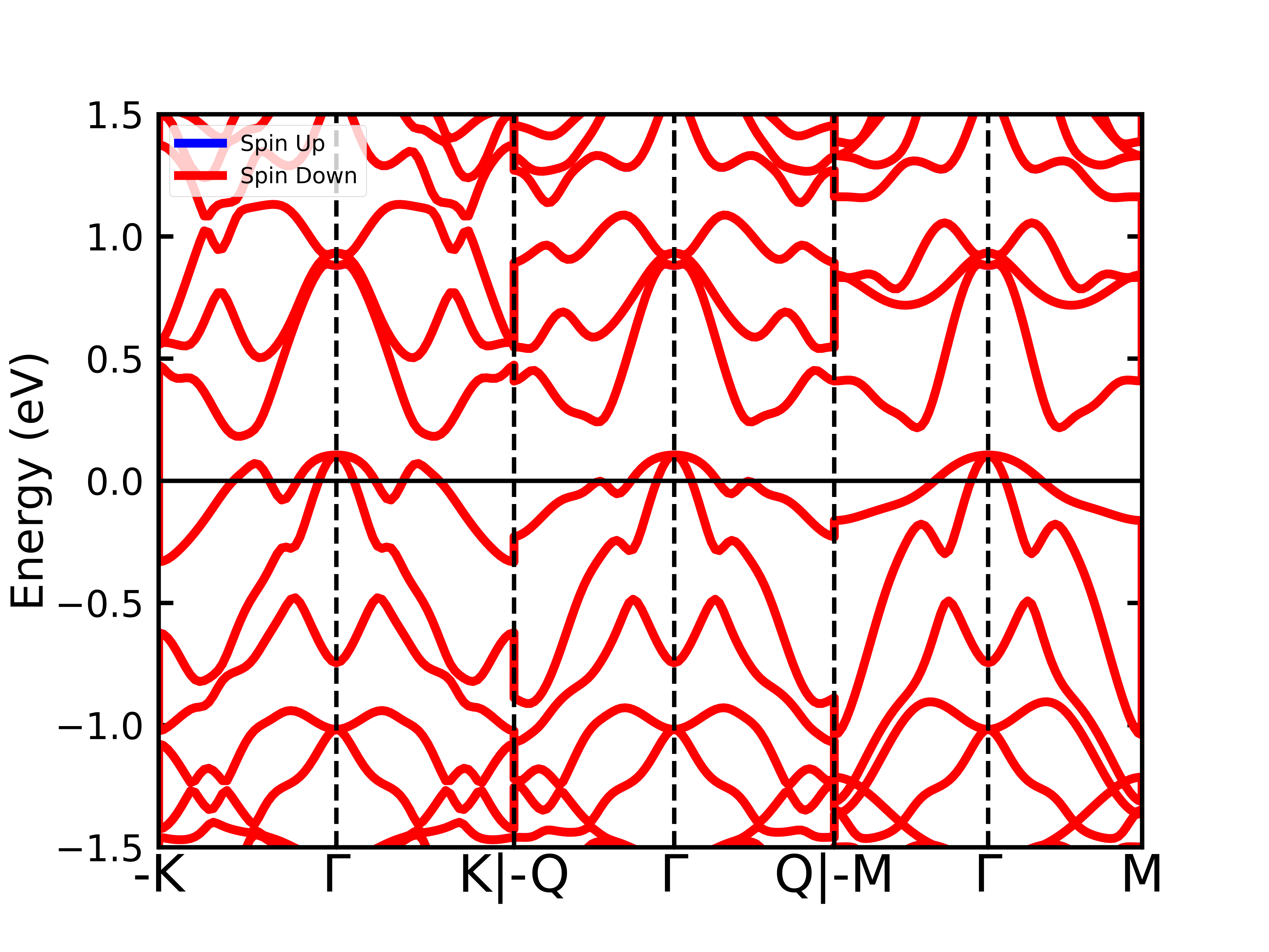}}
    \subfigure[\label{fig:001-SOC}]{\includegraphics[width=0.49\linewidth]{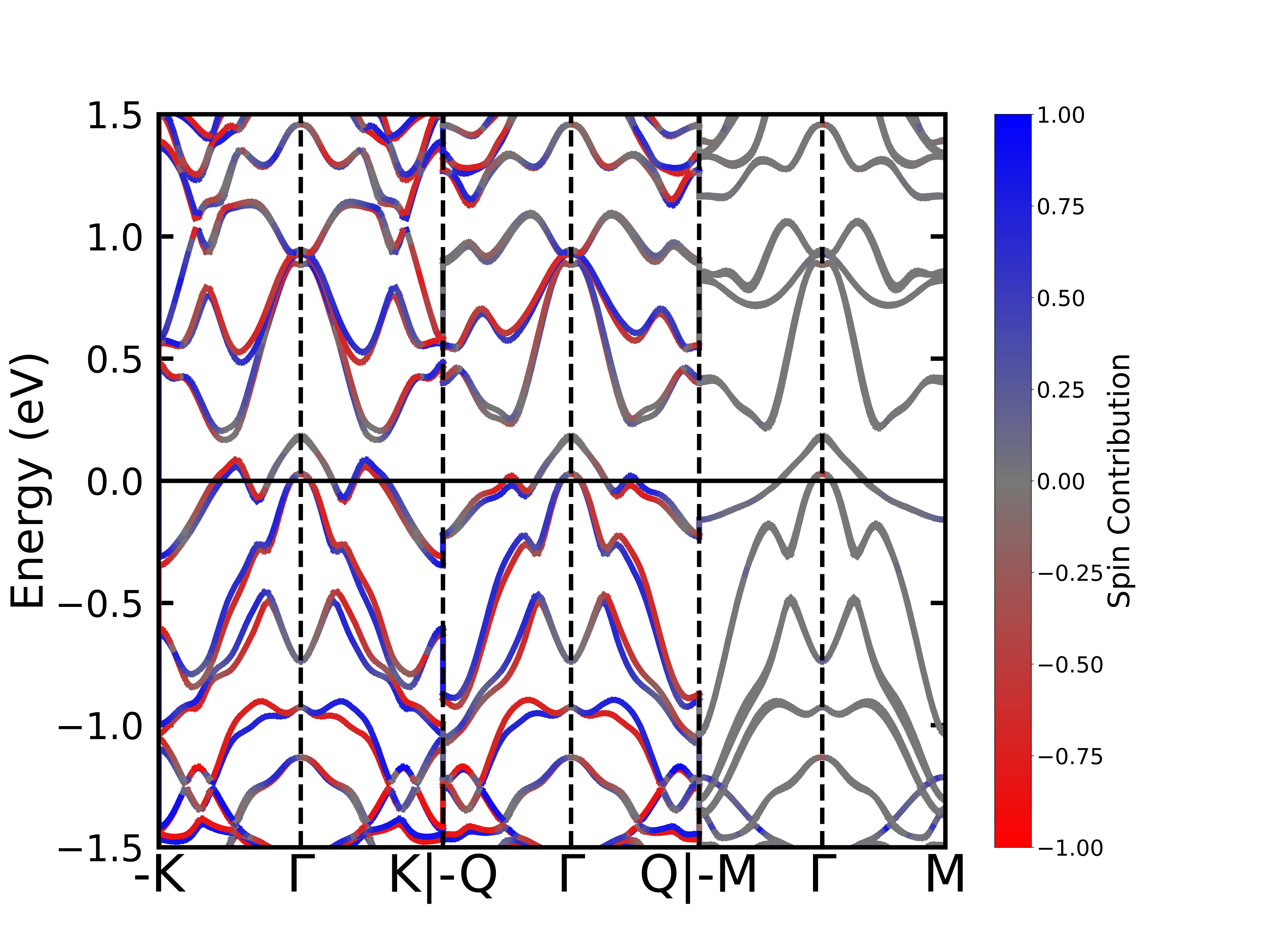}}
     \caption{(a) Structural models of CrSb(0001) slabs: asymmetric Sb$|$Cr termination (left) and symmetric Cr$|$Cr termination (right).
  (b) Band structure of CrSb(0001) slabs with symmetric Cr$|$Cr termination from scalar-relativistic spin-polarized calculations showing spin-degenerate bands, and (c) with SOC. The color bar denotes the spin character of the bands.
  }
    \label{fig:001-bands}
\end{figure*}

Recently, Santhosh \textit{et al.}~\cite{santhosh2025} reported ARPES measurements indicating a large AM splitting ($\sim$700~meV) in CrSb (0001) thin films with thicknesses down to 10 nm.
We therefore investigate whether such behavior survives in the 2D limit by modeling an ultrathin ($\sim$0.9~nm) CrSb(0001) slab. Two surface terminations were considered: an \textit{asymmetric} Sb$|$Cr termination (4 Cr atoms and 4 Sb atoms per unit cell), and a \textit{symmetric} Cr$|$Cr termination (4 Cr atoms and 3 Sb atoms per unit cell) with antiparallel spins on the two surfaces, as shown in Fig.~\ref{fig:001-illustration}. 
To establish which (0001) surface termination is energetically stable, we compared the total energies of the two possible terminations and found that the symmetric Cr$|$Cr termination is more stable by 158\,meV. 
Note that the results presented here are not affected by structural optimization.

In the Sb$|$Cr slab, the under-coordinated surface Cr atom (bonded to a single Sb layer) develops a moment of $+3.8~\mu_\mathrm{B}$, while inner Cr atoms (bulk-like coordination) carry $\pm 2.8~\mu_\mathrm{B}$, producing an uncompensated ferrimagnetic system with a net moment of $1~\mu_\mathrm{B}$ per unit cell. 
Accordingly, this slab shows a conventional ferrimagnetic spin splitting throughout the Brillouin zone, as shown in the Supplementary Material~\cite{suppMat}.

In the Cr$|$Cr slab, surface Cr moments are $\pm3.8~\mu_\mathrm{B}$ and inner Cr moments $\pm2.7~\mu_\mathrm{B}$, yielding zero net magnetization and a collinear A-type AFM state (AFM-1).
The scalar-relativistic band structure is shown in Fig.~\ref{fig:001-noSOC}, indicating that the metallic character is preserved.
Note that we retain the same labels of the symmetry points as in the bulk BZ, Fig.~\ref{fig:BZ}, for clarity.
Along the $\Gamma$–K, $\Gamma$–M, and $\Gamma$–Q lines, the compensated Cr$|$Cr slab is fully spin-degenerate in the scalar-relativistic limit, indicating a collapse of the exchange-driven splitting in the ultrathin limit.
This may be due to the loss of the \Mz\ symmetry; the remaining spin-group symmetries cannot protect the AM splitting.

We also performed additional calculations for significantly thicker CrSb(0001) slabs, with thicknesses up to $\sim 10$ nm. In this regime, the symmetric Cr|Cr termination remains energetically favored, although the energy difference is reduced to about 40 meV per atom. The corresponding band structures are shown in the Supplementary Material~\cite{suppMat}. 
Even for these thicker slabs, the band structures remain symmetric and do not exhibit the altermagnetic splitting characteristic of the bulk Bloch bands.
This reflects the loss of translational symmetry along the surface normal and the reduced slab symmetry, so that $k_z$ is no longer a good quantum number for the finite slab.

Nevertheless, for a sufficiently thick slab, the electronic structure in the interior may still approach that of the bulk. 
By modeling a slab using a tight-binding real-space Hamiltonian derived from the bulk Wannier functions, we have analyzed slabs of varying thickness and calculated the $k_z$-resolved spectral function of interior layers.
 We found that while a the ultrathin slab loses its bulk character, a slab of the thickness presented in experiments (9 nm)\cite{santhosh2025}, shows an interior spectral response that is sufficiently bulk-like for altermagnetic features to become detectable.

Including SOC restores a smaller, momentum-dependent splitting along $\Gamma$–K and $\Gamma$–Q with a magnitude of $\sim$70~meV, while vanishing along $\Gamma$–M, as shown in Fig.~\ref{fig:001-SOC}. 
The orbital moments obtain equal magnitude, so the magnetic order remains fully compensated AFM.
The residual SOC-induced splitting that survives in the ultrathin limit is an order of magnitude smaller than the bulk value, and the (0001) slab may therefore be classified as a weak altermagnetic system~\cite{cheong24}.

\subsection{(100) slab}

\begin{figure}[hbt!]
    \centering
    \subfigure[]{\includegraphics[trim=0 0 0 60,clip,width=0.53\linewidth]{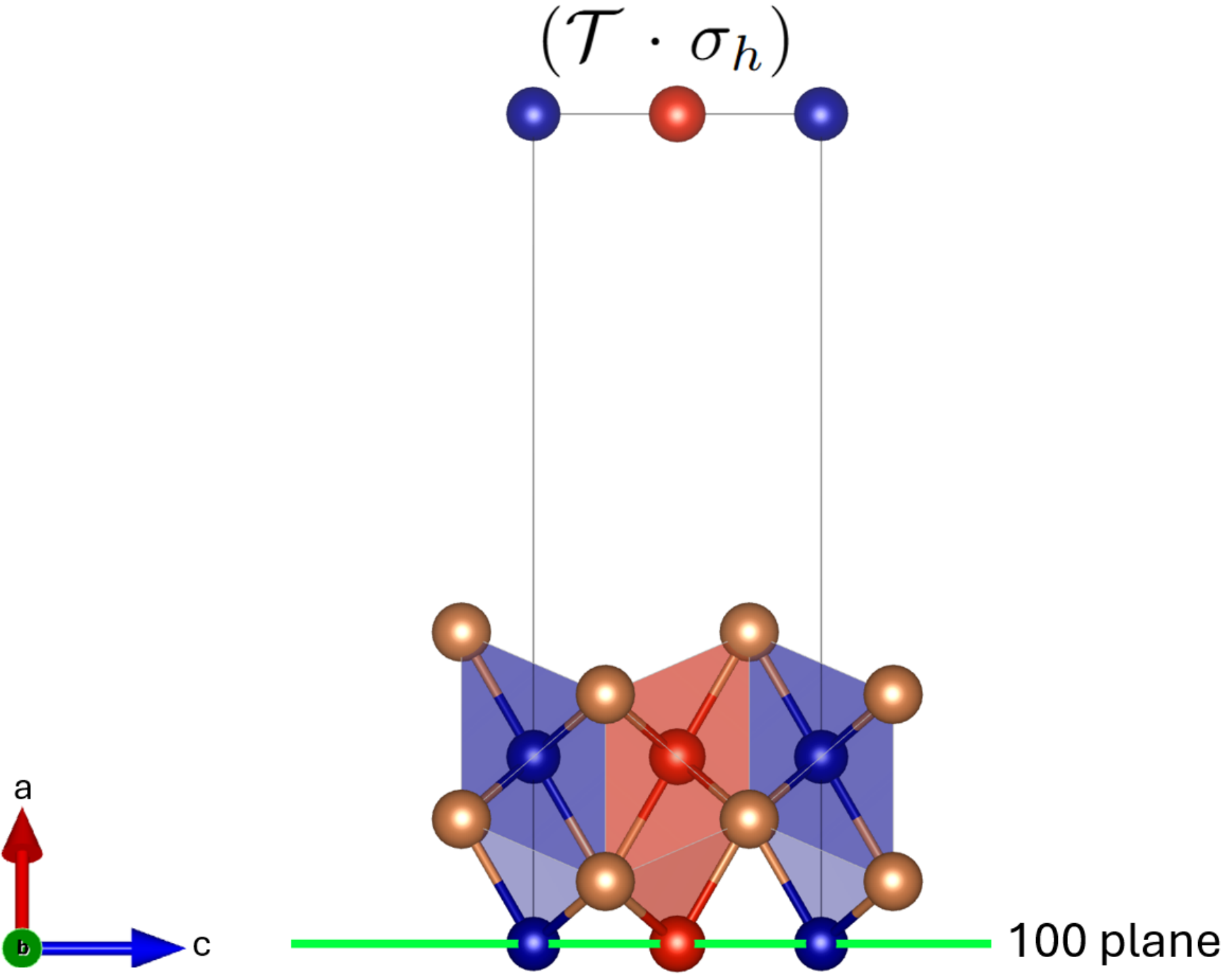}}
    \subfigure[]{\includegraphics[trim=0 80 0 130,clip,width=0.95\linewidth]{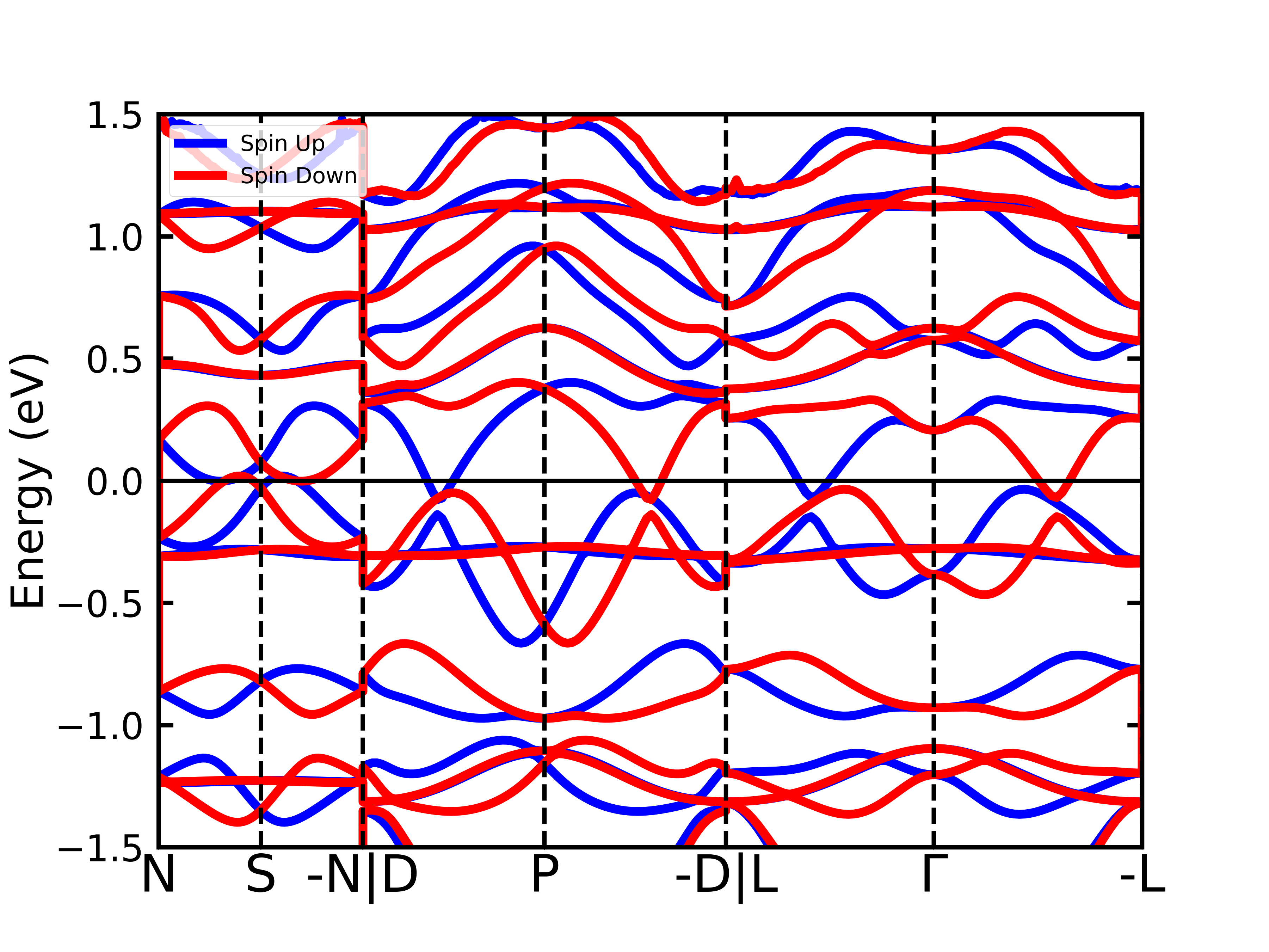}}
    \subfigure[]{\includegraphics[trim=0 80 0 130,clip,width=0.95\linewidth]{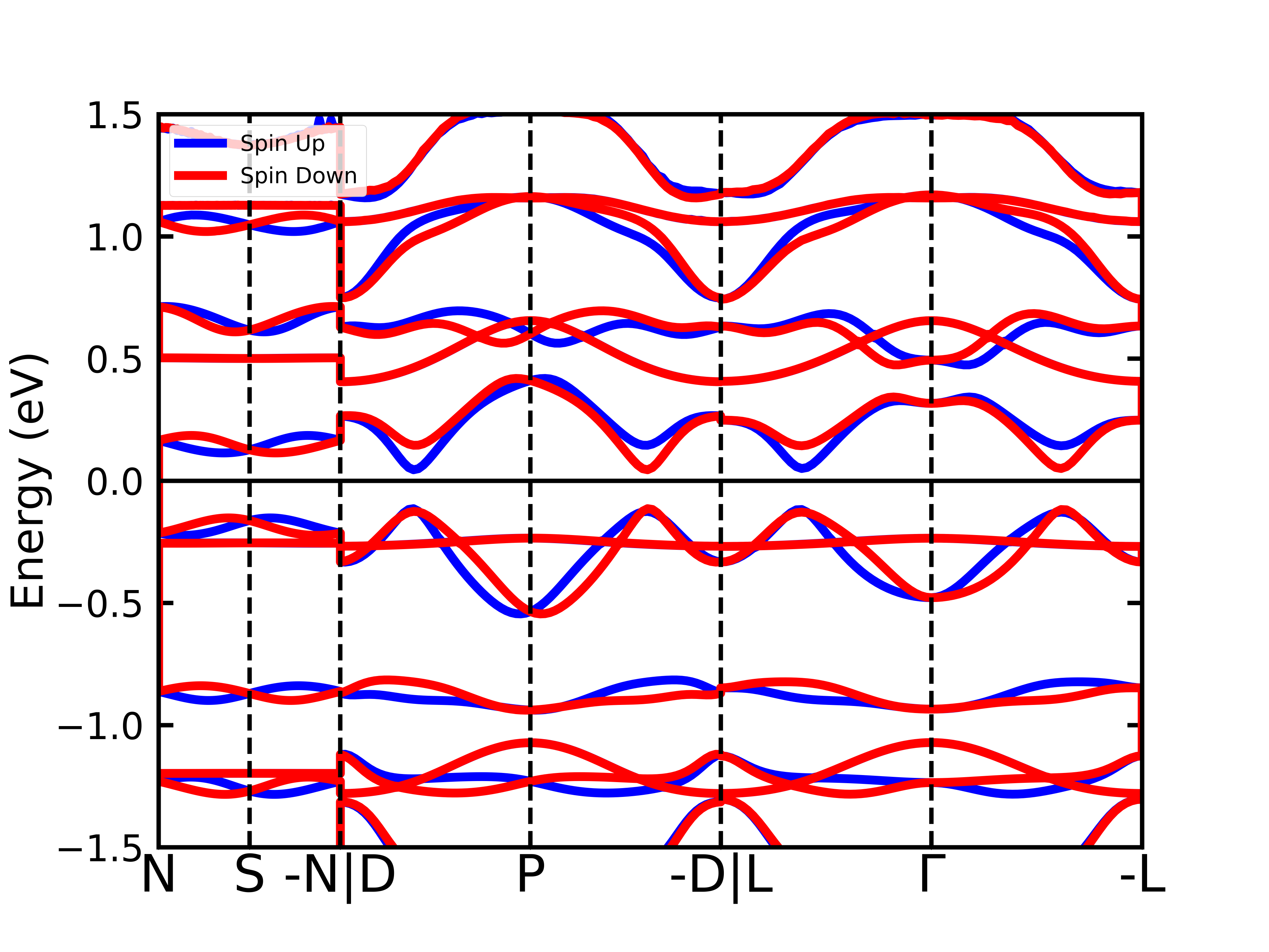}}
    \subfigure[]{\includegraphics[trim=0 80 0 130,clip,width=0.95\linewidth]{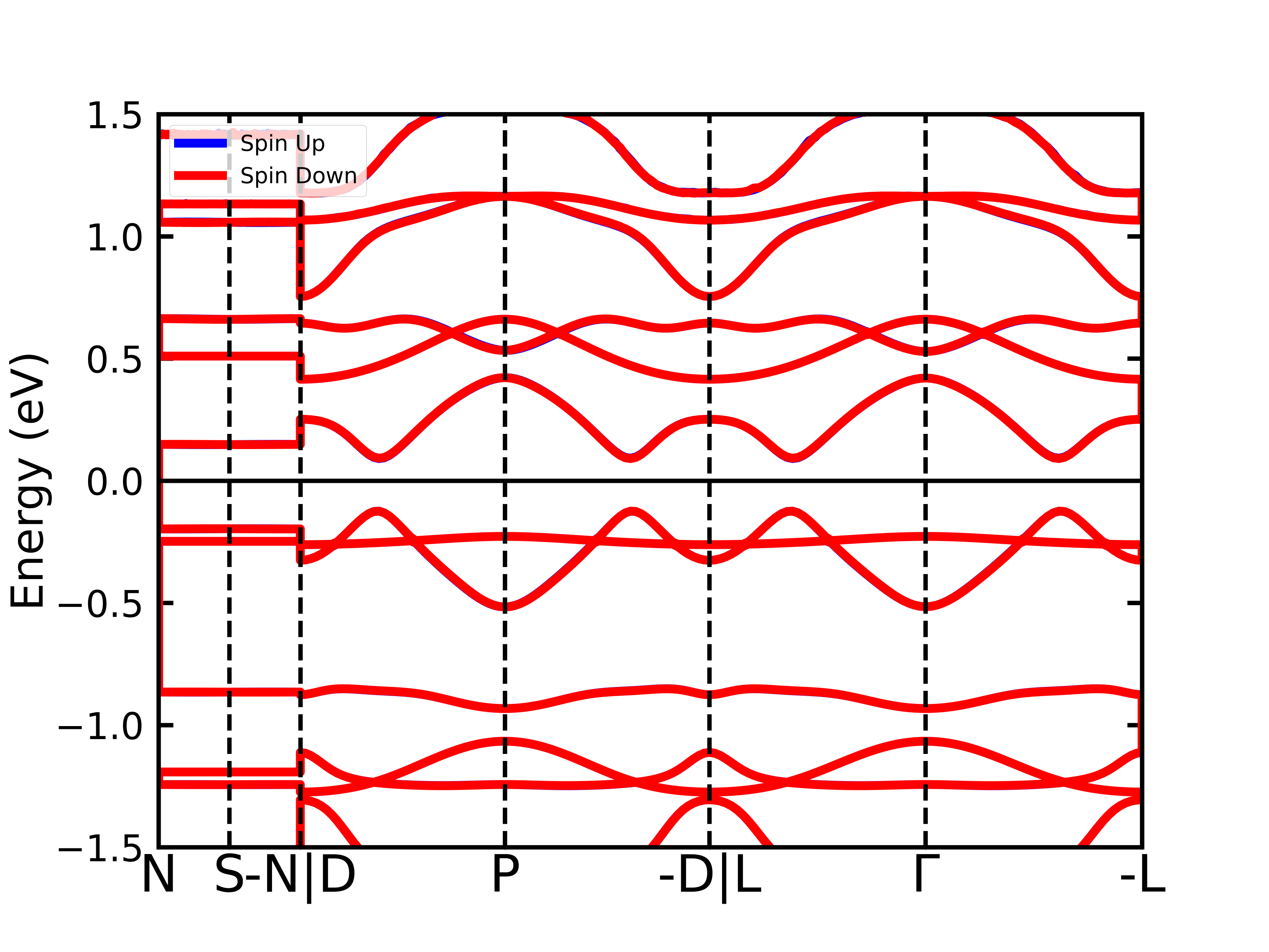}}
    \caption{(a) Illustration of the (100) slab, modeled with a vacuum layer. (b)–(d) Spin‐polarized band structures for vacuum sizes of 3~Å, 5~Å, and 10~Å, respectively.}
    \label{fig:vacuum}
\end{figure}

Figure~\ref{fig:vacuum}(a) shows the (100)-oriented slab model, constructed by cleaving a single Cr$_2$Sb$_2$ layer and introducing vacuum in the $bc$-plane.
The modeled (100) slab is $\sim0.8$ nm thick.
In Figs.~\ref{fig:vacuum}(b)–(d), we show the impact on the spin-polarized band structure as the vacuum layer is progressively increased.
We keep the same path in k-space as for the bulk to show the direct impact on the splitting.
For a vacuum layer of 3 \AA, the AM splitting is retained, but significantly reduced as compared to the bulk bands in Fig.\ \ref{fig:band-1-LS}.

At 5 \AA, the spin splitting is barely visible and at 10 \AA\ the spin-polarized bands are completely spin-degenerate.The bands are also flattened as the translation symmetry in the real-space $x$ direction is broken.
We also examined the band structure projected onto the $k_x=0$ plane of (100)-oriented slab as well as the bulk systems. No altermagnetic spin splitting is found in this cut, as shown in the Supplementary Material \cite{suppMat}. This further confirms that the absence of altermagnetism in the (100) orientation originates from symmetry constraints. Note that the (100) slab also retains its metallicity, as confirmed by the DOS shown in the Supplementary Material \cite{suppMat}.

This suppression is a direct consequence of symmetry breaking: cleaving along (100) destroys the bulk \Csix\ symmetry and retains only the \Mz\ symmetry. 
This single spin–group symmetry appears insufficient to sustain the momentum-dependent exchange splitting characteristic of altermagnetism, resulting in spin-degenerate bands in the ultrathin limit. 
Even a symmetrically terminated Cr$_4$Sb$_2$ slab with $C_{2h}$ crystal symmetry and $S_{2}$ magnetic point group remains spin-degenerate, as shown in the Supplementary Material \cite{suppMat}.
The same results are obtained with SOC included \cite{suppMat}. Note that the conclusions for the (100) slab are unchanged by structural optimization.

In addition, we also computed a substantially thicker (100) slab with a total thickness of $\sim 10$\,nm. The resulting scalar-relativistic band structure, as shown in the Supplementary Material \cite{suppMat}, remains fully spin-degenerate, with no reemergence of the exchange-driven splitting. This is because the non-symmorphic symmetries that protect the bulk splitting are intrinsically broken by the (100) cleavage, and are not reinstated by adding more layers in calculations.

Thus, the (100) oriented slabs do not sustain AM splitting in the ultrathin limit: the exchange-driven splitting collapses due to the loss of bulk $C_{6}$ rotational symmetry, and SOC fails to reintroduce any momentum-dependent splitting within our resolution.  
Based on these results, we suggest that the ARPES measurements on films of 30 nm thickness in Ref.\ \cite{Reimers2024}, which reported AM splitting along the ($-D$)--$P$--$D$ path, may instead reflect the electronic structure of bulk-like states rather than an intrinsic property of ultrathin films. This is analogous to the (001) surface, as can be confirmed from the analyses based on the tight-binding real-space Hamiltonian \cite{suppMat}.

\subsection{(110) slab}\label{sec:slab110}

\begin{figure}[hbt!]
    \centering
    \subfigure[\label{fig:slab-110-geometry}]{
 \includegraphics[trim=0 0 0 0,clip,width=0.8\linewidth]{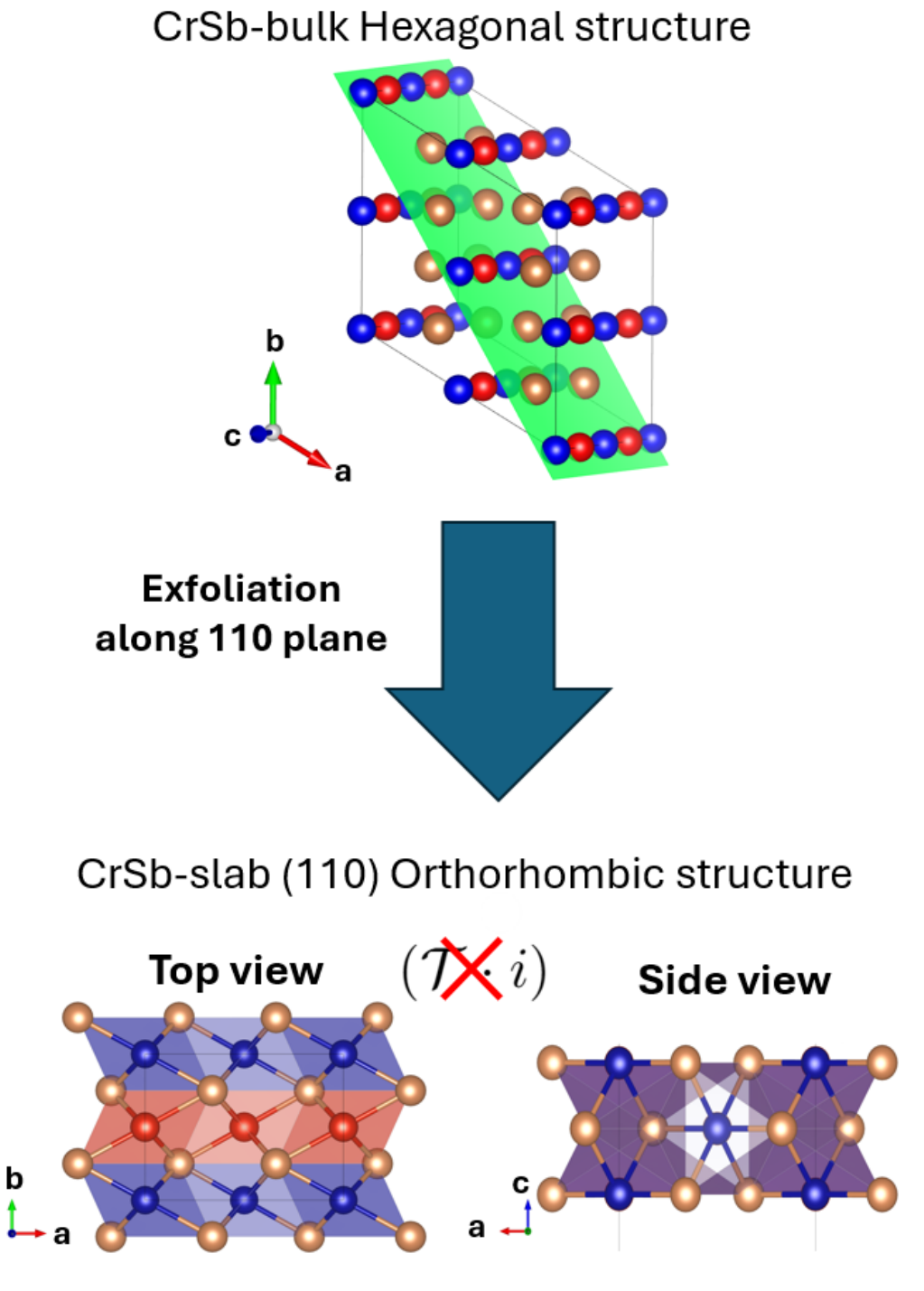}}
   \subfigure[\label{fig:slab-110-BZ}]{\includegraphics[width=0.4\linewidth]{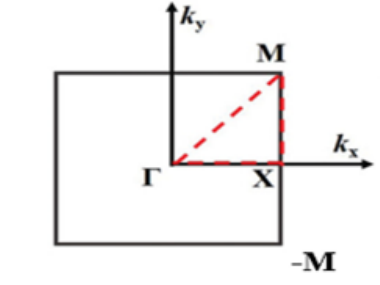}}
    \subfigure[\label{fig:slab-110-bandstructure}]{\includegraphics[trim=0 90 0 90,clip,width=0.5\textwidth]{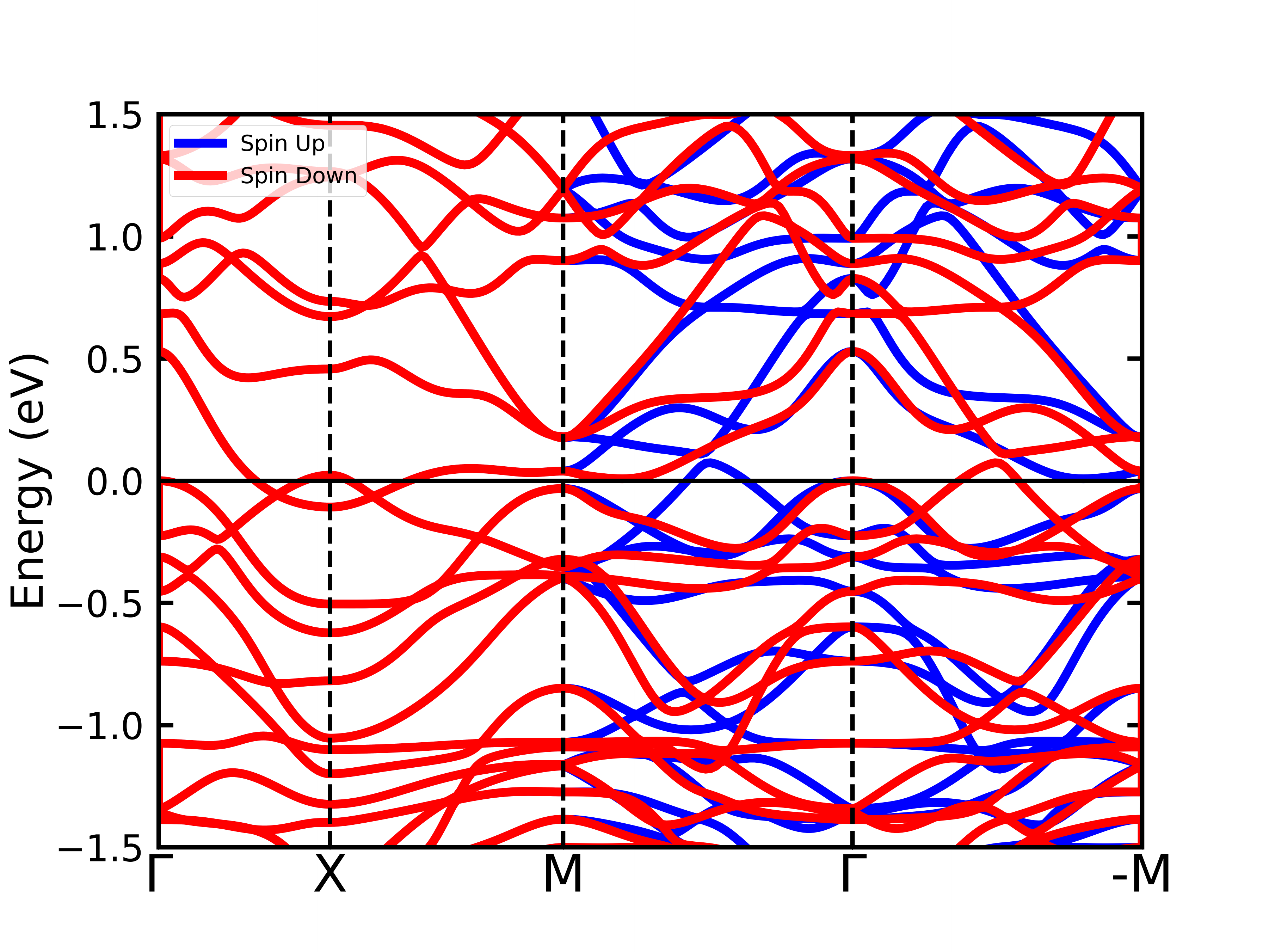}}
       \caption{(a) (110) exfoliation plane in the hexagonal CrSb and the orthorhombic slab. (b) First Brillouin zone (BZ) of the orthorhombic slab. (c) Spin-polarized band structure of slab without SOC. The spin-up and spin-down states are indicated by the blue red lines, respectively.
    }
    \label{fig:slab-110}
\end{figure}

The (110) slab, shown in Fig.~\ref{fig:slab-110-geometry}, has orthorhombic symmetry, and the optimized in-plane lattice parameters $a = 7.16 \, \text{\AA}$ and $b = 5.26 \, \text{\AA}$.
Energy comparisons of different magnetic configurations confirm that the bulk AFM ground state is retained in this orientation (Table~\ref{tab:Ediff-slab}). 
Furthermore, the (110) slab exhibits an in-plane magnetic easy axis, with a magnetic anisotropy energy of -0.17~meV/atom.  

The orthorhombic Brillouin zone [Fig.~\ref{fig:slab-110-BZ}] contains the $(-M)$–$\Gamma$–$M$ line, along which a robust $\sim$0.4~eV spin splitting persists while the electronic structure remains metallic, as shown in Fig.~\ref{fig:slab-110-bandstructure}.

The magnitude of this splitting, comparable to the bulk, indicates that the exchange mechanism is fully preserved in this cut. 
Including SOC leaves the spin splitting almost unchanged, with only minor variations below 5~meV, as shown in the Supplementary Material \cite{suppMat}. This confirms that the AM splitting is primarily exchange-driven.

Like the (100) slab, the (110) orientation breaks the bulk $C_{6}$ rotation symmetry, but it preserves both the \Mz\ and \Mxy\ spin-group symmetries of the bulk. These spin-group symmetries seem to preserve the AM splitting in parts of the Brillouin zone.

\begin{table}
\centering
\caption{Energy differences ($E_{\mathrm{diff}}$) relative to AFM-1 and local Cr magnetic moments for the CrSb(110) slab.}
\label{tab:Ediff-slab}
\begin{ruledtabular}
\begin{tabular}{lcccc} 
Config. & FM & AFM-1 & AFM-2 & AFM-3\\[0.5ex]
$E_{\mathrm{diff}}$ (meV/atom) & 21.01 & 0.00 & 42.82 & 103.36\\
$M_{\mathrm{Cr}}$ ($\mu_B$) & 3.36 & 3.38 & 3.40 & 3.38\\ 
\end{tabular}
\end{ruledtabular}
\end{table}

\subsection{Magnetic Exchange Interactions}

\begin{figure} 
    \centering
    \subfigure[]{\includegraphics[width=1\linewidth]{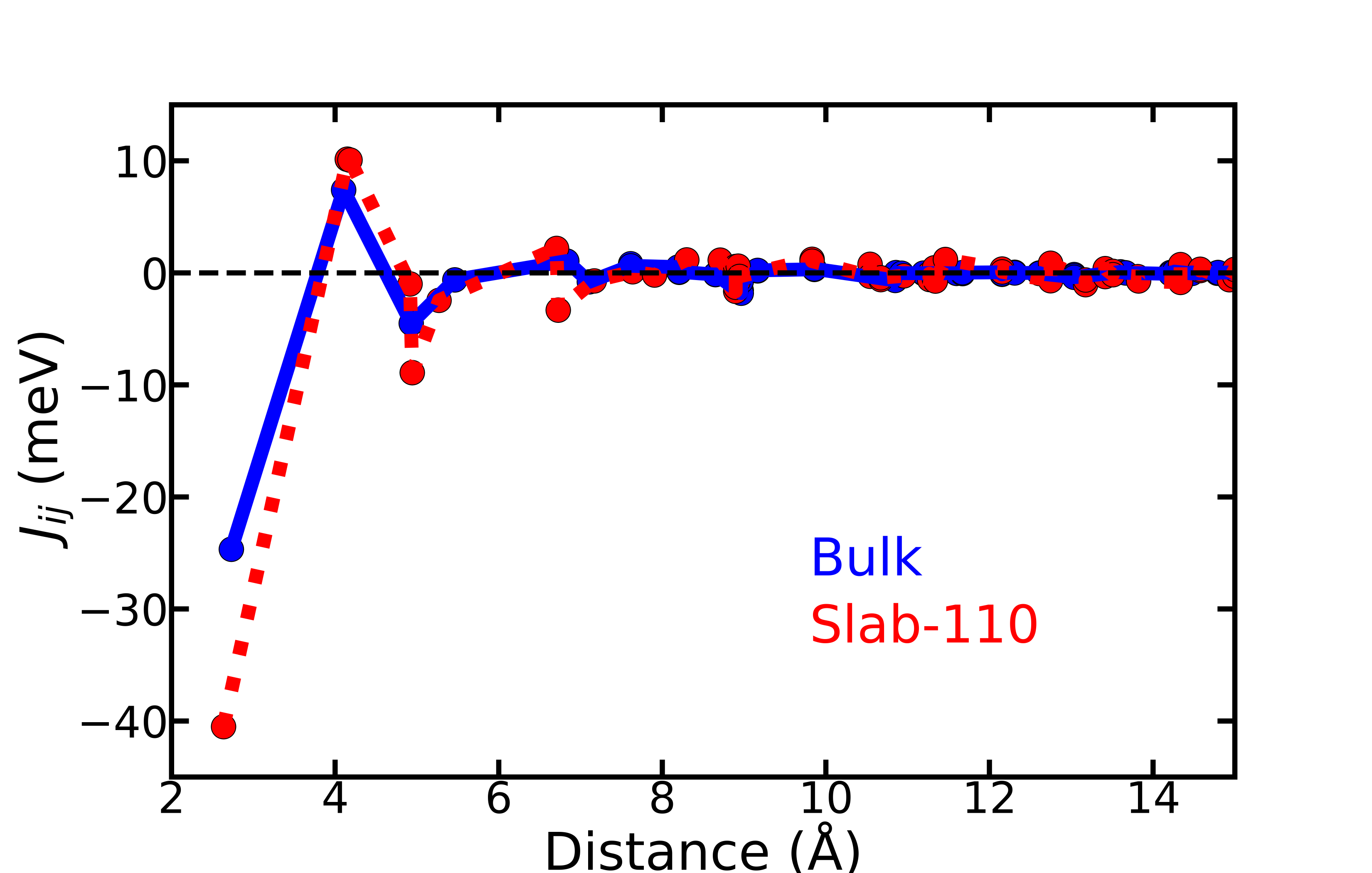}}
    \subfigure[]{\includegraphics[width=1\linewidth]{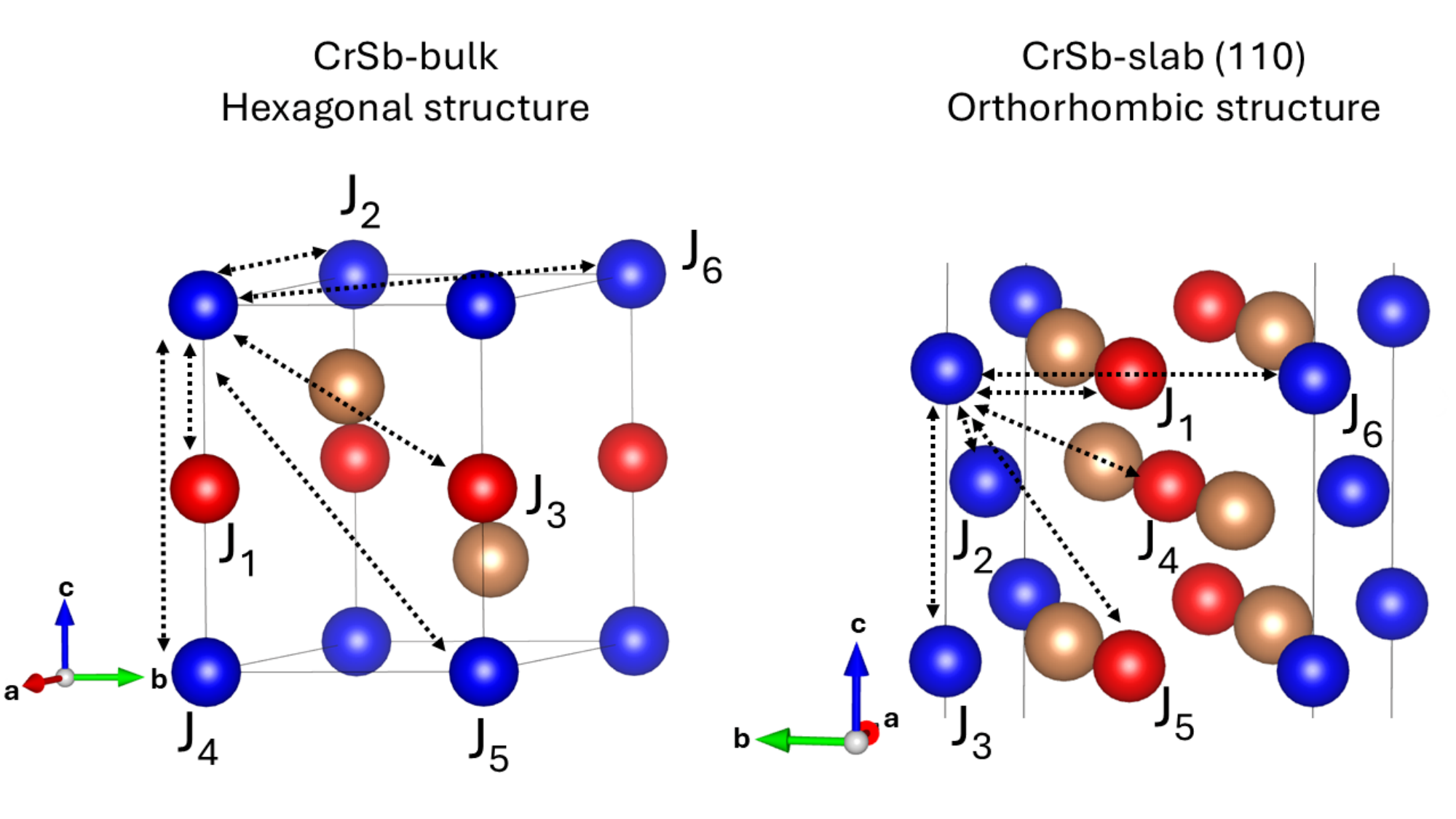}}
    \caption{(a) Calculated isotropic Heisenberg exchange parameters $J_{ij}$ as a function of interatomic distance $d_{ij}$, obtained via the magnetic force theorem. (b) Schematic depiction of the $J_i$ exchange pathways.}
    \label{fig:exchange}
\end{figure}

In order to further assess the impact of reduced dimensionality in the (110) slab on the magnetic properties of the CrSb system, we have computed the isotropic exchange parameters of the Heisenberg Hamiltonian:
\begin{equation}\label{eq:heisenberg_hamiltonian}
    H = -\sum_{i\neq j} J_{ij} \, \mathbf{\hat{e}}_i \cdot \mathbf{\hat{e}}_j \, ,
\end{equation}
where $\mathbf{\hat{e}}_i$ is the unit vector along the local moment at site $i$. A positive (negative) $J_{ij}$ denotes a FM (AFM) interaction. The calculated $J_i$ values for the bulk and (110) slab are shown in Fig.~\ref{fig:exchange}, with the dominant interactions summarized in Table~\ref{tab:exchange_constants}.

For the bulk system, the first nearest-neighbor (NN) interaction $J_1$ is strongly AFM ($-24.66$~meV), coupling $\mathrm{Cr}^{\uparrow}$ to $\mathrm{Cr}^{\downarrow}$ along the $\mathbf{c}$ axis. The second NN interaction $J_2 = 7.40$~meV is FM and connects parallel-spin $\mathrm{Cr}$ sites along $\mathbf{a}$. The third NN is again AFM ($J_3=-4.50$~meV). The fourth and sixth NNs ($J_4=-0.62$~meV, $J_6=-0.80$~meV) are AFM despite linking parallel-spin Cr atoms, while $J_5 = 1.05$~meV is FM. 
The alternating signs reflect the competing exchange pathways characteristic of an itinerant system.

For the (110) slab, $J_1$ is even more strongly AFM ($-40.50$~meV), reflecting enhanced in-plane Cr–Cr exchange in the reduced dimensionality. 
The next two NNs are both FM ($J_2=10.14$~meV, $J_3=10.07$~meV), while $J_4=-0.98$~meV, $J_5=-8.90$~meV, and $J_6=-2.45$~meV are AFM. 
Notably, in the (110) slab all interactions are compatible with the AFM ground state. This suggests that dimensional confinement in this orientation selectively amplifies the dominant exchange while suppressing competing terms.

These results reinforce the robustness of the (110) orientation discussed in Sec.~\ref{sec:slab110}: its exchange topology preserves strong AFM coupling down to the ultrathin limit, enabling the survival of large exchange-driven spin splitting.

To quantify the associated ordering energy scale,  we evaluate the effective exchange parameter obtained from the shell-averaged couplings
\begin{equation}
J_0 = \sum_i z_i J_i ,
\label{eq:J0}
\end{equation}
where $z_i$ is the coordination number of the $i$-th exchange shell and $J_i$ are the isotropic exchange constants listed in Table~\ref{tab:exchange_constants}. Using these coordination-weighted couplings, we obtain $J_0^{\mathrm{bulk}} = -52.36$~meV for bulk CrSb and $J_0^{\mathrm{slab}} = -65.2$~meV for the (110) slab. Note that the local Cr magnetic moments are also comparable, with
$m_{\mathrm{bulk}} = 2.59\,\mu_{\mathrm B}$ and $m_{\mathrm{slab}} = 3.08\,\mu_{\mathrm B}$.

The fact that $J_0^{\mathrm{slab}}$ is even slightly larger in magnitude than $J_0^{\mathrm{bulk}}$ indicates that the dominant exchange energy scale is preserved, and in fact mildly enhanced, in the (110) orientation. Since the mean-field N\'eel temperature scales with both $m$ and $|J_0|$, this comparison implies that the ordering temperature of the slab should remain of the same order as that of the bulk. These results indicate that the (110) orientation maintains a high antiferromagnetic energy scale. However, an accurate determination of transition temperatures necessitates going beyond mean-field, including non-isotropic terms of the Hamiltonian, and accounting for the renormalization of the exchange interactions due to thermal effects.

\begin{table}
\caption{Isotropic exchange constants $J_i$ and distances $d_i$ between the $i$-th nearest neighbors for bulk and (110) slab $\mathrm{CrSb}$. Positive (negative) $J_i$ indicates FM (AFM) coupling.} 
 \begin{ruledtabular}
     \begin{tabular}{l c c c c c c } 
    $i$ & 1 & 2 & 3 & 4 & 5 & 6\\
         \hline
    Bulk & & & & & & \\ 
    $J_i$ (meV) & -24.66 & 7.40 & -4.50 & -0.62 & 1.05 & -0.80\\ 
    $d_i$ (\AA) & 2.73 & 4.10 & 4.92 & 5.46 & 6.83 & 7.10\\ 
    \hline
    (110) slab & & & & & & \\ 
    $J_i$ (meV) & -40.50 & 10.14 & 10.07 & -0.98 & -8.90 & -2.45\\ 
    $d_i$ (\AA) & 2.63 & 4.15 & 4.18 & 4.91 & 4.94 & 5.27\\ 
\end{tabular}
\label{tab:exchange_constants}
 \end{ruledtabular}
 \end{table}

\section{Summary and Conclusions}

We have investigated the dependence of altermagnetic spin splitting on crystallographic orientation in ultrathin $\mathrm{CrSb}$ slabs using first-principles calculations. Bulk $\mathrm{CrSb}$ shows a large exchange-driven altermagnetic spin splitting of order 0.6--1~eV, as established by DFT calculations and recent ARPES measurements~\cite{Reimers2024, santhosh2025}. However, upon dimensional reduction, the persistence of altermagnetism is found to be strongly orientation dependent.

In (0001)-oriented ultrathin slabs, we find that the exchange-driven splitting collapses, which may be explained by the loss of the \Mz\ spin-group symmetry. The remaining symmetries do not sustain the altermagnetic splitting. Including SOC restores a small splitting of $\sim 70$~meV, that is qualitatively distinct from the bulk exchange-driven altermagnetic mechanism.

The (100) oriented ultrathin slab behaves differently: cleaving reduces the symmetry to \Mz\ only, and the \Csix\ spin-group symmetry is lost. 
But this is insufficient to sustain momentum-dependent exchange splitting, as the ultrathin (100) slab becomes fully spin degenerate, even when SOC is included.

In contrast, the (110) orientation preserves both the \Mxy\ and \Mz\ symmetries, although \Csix\ is broken. In this case, a large exchange-driven splitting of $\sim 0.4$~eV survives along the $(-M)$–$\Gamma$–$M$ direction, comparable to the bulk value. Moreover, the magnetic exchange interactions in this geometry remain robust and are even enhanced relative to the bulk, underscoring the stability of the altermagnetic energy scale.

These results establish the (110) orientation as a particularly well-defined and robust platform for realizing exchange-driven altermagnetism even for ultrathin films.
Recent experimental reports demonstrating the epitaxial growth of CrSb(110) films~\cite{aota2025} further reinforce this conclusion and open a clear experimental pathway for observing strong altermagnetism in ultrathin CrSb films.

\begin{acknowledgments}
We acknowledge financial support from Olle Engkvists Stiftelse (Project No.~207-0582), the Swedish e-Science Research
Centre (SeRC), and Flair.
The computations were enabled by resources provided by the National Academic Infrastructure for Supercomputing in Sweden (NAISS), partially funded by the Swedish Research Council through grant agreement no. 2022-06725, and by resources provided by the National Supercomputer Centre (NSC), funded by Link\"oping University.

\end{acknowledgments}


\begin{thebibliography}{}
\bibitem{Mazin2022} I. Mazin, P. Editors, Altermagnetism—a new punch line of fundamental magnetism, APS News, 040002 (2022).

\bibitem{Smejkal2022} L. \v{S}mejkal, J. Sinova, and T. Jungwirth, Emerging research landscape of altermagnetism, Phys. Rev. X \textbf{12}, 040501 (2022).

\bibitem{Lee2024} S. Lee, S. Lee, S. Jung, J. Jung, D. Kim, Y. Lee \textit{et al.}, Broken Kramers degeneracy in altermagnetic MnTe, Phys. Rev. Lett. \textbf{132}, 036702 (2024).

\bibitem{Krempasky2024} J. Krempaský, L. Šmejkal, S. D’Souza, M. Hajlaoui, G. Springholz, K. Uhlířová \textit{et al.}, Altermagnetic lifting of Kramers spin degeneracy, Nature \textbf{626}, 517--522 (2024).

\bibitem{Feng2022} Z. Feng, X. Zhou, L. Šmejkal, L. Wu, Z. Zhu, H. Guo \textit{et al.}, An anomalous Hall effect in altermagnetic ruthenium dioxide, Nature Electronics \textbf{5}, 735--743 (2022).

\bibitem{Smejkal2022a} L. \ifmmode \check{S}\else \v{S}\fi{}mejkal, J. Sinova, and T. Jungwirth, Beyond Conventional Ferromagnetism and Antiferromagnetism: A Phase with Nonrelativistic Spin and Crystal Rotation Symmetry, Phys. Rev. X \textbf{12}, 031042 (2022).

\bibitem{Brinkman1966} W. F. Brinkman, R. J. Elliott, and R. E. Peierls, Theory of spin-space groups, Proc. R. Soc. Lond. A \textbf{294}, 343--358 (1966).

\bibitem{Fedchenko2024} O. Fedchenko, J. Minár, A. Akashdeep, S. D’Souza, D. Vasilyev, O. Tkach \textit{et al.}, Observation of time-reversal symmetry breaking in the band structure of altermagnetic RuO$_2$, Sci. Adv. \textbf{10}, eadj4883 (2024).

\bibitem{Guo2024} Y. Guo, J. Zhang, Z. Zhu, Y. Jiang, L. Jiang, C. Wu \textit{et al.}, Direct and Inverse Spin Splitting Effects in Altermagnetic RuO$_2$, Adv. Sci., 2400967 (2024).

\bibitem{Osumi2024} T. Osumi, S. Souma, T. Aoyama, K. Yamauchi, A. Honma, K. Nakayama \textit{et al.}, Observation of a giant band splitting in altermagnetic MnTe, Phys. Rev. B \textbf{109}, 115102 (2024).

\bibitem{Hariki2024} A. Hariki, D. A. Dal, O. Amin, T. Yamaguchi, A. Badura, D. Kriegner \textit{et al.}, X-ray magnetic circular dichroism in altermagnetic $\alpha$-MnTe, Phys. Rev. Lett. \textbf{132}, 176701 (2024).

\bibitem{takei63} W. J. Takei, D. E. Cox, and G. Shirane, Magnetic Structures in the MnSb-CrSb System, Phys. Rev. \textbf{129}, 2008--2018 (1963).

\bibitem{Yang2024} G. Yang, Z. Li, S. Yang, J. Li, H. Zheng, W. Zhu, Z. Pan, Y. Xu, S. Cao, W. Zhao \textit{et al.}, Three-dimensional mapping of the altermagnetic spin splitting in CrSb, Nat. Commun. \textbf{16}, 1442 (2025).

\bibitem{Li2024} C. Li, M. Hu, Z. Li, Y. Wang, W. Chen, B. Thiagarajan, M. Leandersson, C. Polley, T. Kim, H. Liu \textit{et al.}, Topological Weyl altermagnetism in CrSb, Commun. Phys. \textbf{8}, 311 (2025).

\bibitem{Ding2024} J. Ding, Z. Jiang, X. Chen, Z. Tao, Z. Liu, T. Li, J. Liu, J. Sun, J. Cheng, J. Liu, Y. Yang, R. Zhang, L. Deng, W. Jing, Y. Huang, Y. Shi, M. Ye, S. Qiao, Y. Wang, Y. Guo, D. Feng, and D. Shen, Large Band Splitting in $g$-Wave Altermagnet CrSb, Phys. Rev. Lett. \textbf{133}, 206401 (2024).

\bibitem{Lu2024} W. Lu, S. Feng, Y. Wang, D. Chen, Z. Lin, X. Liang, S. Liu, W. Feng, K. Yamagami, J. Liu \textit{et al.}, Signature of Topological Surface Bands in Altermagnetic Weyl Semimetal CrSb (2025).

\bibitem{Reimers2024} S. Reimers, L. Odenbreit, L. \v{S}mejkal, V. Strocov, P. Constantinou, A. Hellenes \textit{et al.}, Direct observation of altermagnetic band splitting in CrSb thin films, Nat. Commun. \textbf{15}, 2116 (2024).

\bibitem{rai2025direction} B. Rai, K. Patra, S. Bera, S. Kalimuddin, K. Deb, M. Mondal, P. Mahadevan, and N. Kumar, Direction-Dependent Conduction Polarity in Altermagnetic CrSb, Adv. Sci., 2502226 (2025).

\bibitem{Zhou2025} Z. Zhou, X. Cheng, M. Hu, R. Chu, H. Bai, L. Han, J. Liu, F. Pan, and C. Song, Manipulation of the altermagnetic order in CrSb via crystal symmetry, Nature \textbf{638}, 645--650 (2025).

\bibitem{zhang2025chiral} Y. Zhang, X. Ni, K. Chen, and K. Cao, Chiral magnon splitting in altermagnetic CrSb from first principles, Phys. Rev. B \textbf{111}, 174451 (2025).

\bibitem{yu2025neel} T. Yu, I. Shahid, P. Liu, D. Shao, X. Chen, and Y. Sun, Néel vector-dependent anomalous transport in altermagnetic metal CrSb, npj Quantum Mater. \textbf{10}, 1--7 (2025).

\bibitem{Urata2024} T. Urata, W. Hattori, and H. Ikuta, High mobility charge transport in a multicarrier altermagnet CrSb, Phys. Rev. Mater. \textbf{8}, 084412 (2024).

\bibitem{venkatraman1990} M. Venkatraman and J. Neumann, The Cr-Sb (chromium-antimony) system, J. Phase Equilib. \textbf{11}, 435--440 (1990).

\bibitem{Sim2025} J. Sim, C. Kang, and S. Je, Leveraging Strain-Induced Staggered Dzyaloshinskii-Moriya Interaction for Altermagnetic Néel Vector Control, Physica B, 417475 (2025).

\bibitem{santhosh2025} S. Santhosh, P. Corbae, W. J. Yánez-Parreño, S. Ghosh, C. J. Jensen, A. V. Fedorov, M. Hashimoto, D. Lu, J. A. Borchers, A. J. Grutter \textit{et al.}, Altermagnetic band splitting in 10 nm epitaxial CrSb thin films, Adv. Mater., e08977 (2025).

\bibitem{aota2025} S. Aota and M. Tanaka, Epitaxial growth and transport properties of a metallic altermagnet CrSb on a GaAs (001) substrate, Phys. Rev. Mater. \textbf{9}, 074410 (2025).

\bibitem{Kresse1999} G. Kresse and D. Joubert, From ultrasoft pseudopotentials to the projector augmented-wave method, Phys. Rev. B \textbf{59}, 1758 (1999).

\bibitem{Blochl1994} P. E. Blöchl, Projector augmented-wave method, Phys. Rev. B \textbf{50}, 17953 (1994).

\bibitem{Perdew1996} J. P. Perdew, K. Burke, and M. Ernzerhof, Generalized gradient approximation made simple, Phys. Rev. Lett. \textbf{77}, 3865 (1996).

\bibitem{Pizzi2020} G. Pizzi, V. Vitale, R. Arita, S. Blügel, F. Freimuth, G. Géranton, M. Gibertini, D. Gresch, C. Johnson, T. Koretsune \textit{et al.}, Wannier90 as a community code: new features and applications, J. Phys.: Condens. Matter \textbf{32}, 165902 (2020).

\bibitem{He2021} X. He, N. Helbig, M. Verstraete, and E. Bousquet, TB2J: A python package for computing magnetic interaction parameters, Comput. Phys. Commun. \textbf{264}, 107938 (2021).

\bibitem{snow52} A. I. Snow, Neutron Diffraction Investigation of the Atomic Magnetic Moment Orientation in the Antiferromagnetic Compound CrSb, Phys. Rev. \textbf{85}, 365--365 (1952).

\bibitem{suppMat} See Supplemental Material at [URL] for detailed thickness-dependent resolved spectral and electronic structure data for (0001) and (100) slabs.

\bibitem{cheong24} S. Cheong and F. Huang, Altermagnetism classification, npj Quantum Mater. \textbf{10}, 38 (2025).
\end{thebibliography}
\end{document}